\documentclass[a4paper,12pt]{article}

\usepackage[sumlimits,intlimits,namelimits]{amsmath}
\usepackage{amssymb}
\usepackage{epstopdf}

\usepackage[english]{babel}

\usepackage{graphicx}
\usepackage{subfigure}
\usepackage{epstopdf}
\usepackage{multirow, rotating}

\newcommand{\lsim}{
\mathrel{\hbox{\rlap{\hbox{\lower4pt\hbox{$\sim$}}}\hbox{$<$}}}}

\newcommand{\gsim}{
\mathrel{\hbox{\rlap{\hbox{\lower4pt\hbox{$\sim$}}}\hbox{$>$}}}}

\newcommand{\MJpsi}{\ensuremath{m_{J/\psi}^2}}
\newcommand{\Bphidec}{\ensuremath{B_s \to \phi}}
\newcommand{\BsKdec}{\ensuremath{B_s \to \bar K^\ast}}

\def\D0{D\O }
\begin{document}
\begin{titlepage}
\vspace*{-0.5truecm}

\begin{flushright}
CERN-PH-TH/2008-167\\
SI-HEP-2008-15
\end{flushright}

\vspace*{-0.8truecm}

\begin{center}
\boldmath
{\Large{\bf \mbox{Precision Physics with $B^0_s\to J/\psi \phi$ at the LHC}:\\[0.2cm]
The Quest for New Physics}}
\unboldmath
\end{center}

\vspace{0.4truecm}

\begin{center}
{\sc Sven Faller,${}^{a,b}$ Robert Fleischer${}^{a,c}$} and
{\sc Thomas Mannel${}^{a,b}$}

\vspace{0.4truecm}

${}^a$ {\sl Theory Division, Department of Physics, CERN, CH-1211 Geneva 23,
Switzerland}

\vspace{0.2truecm}

${}^b$ {\sl Theoretische Physik 1, 
Fachbereich Physik, Universit\"at Siegen, \\
D-57068 Siegen, Germany}

\vspace{0.2truecm}

${}^c$ {\sl Dipartimento di Fisica, Universit\`a di Roma ``La Sapienza", \\
I-00185 Roma, Italy }

\end{center}

\vspace{0.6cm}
\begin{abstract}
\vspace{0.2cm}\noindent
CP-violating effects in the time-dependent angular distribution of the
$B^0_s\to J/\psi[\to\ell^+\ell^-]\phi[\to K^+K^-]$ decay products play a 
key r\^ole for the search of new physics. The hadronic Standard-Model
uncertainties are related to doubly Cabibbo-suppressed penguin 
contributions and are usually assumed to be negligibly small. In view of 
recent results from the Tevatron and the quickly approaching start of the 
data taking at the LHC, we have a critical look at the impact of these terms, 
which could be enhanced through long-distance QCD phenomena, and 
explore the associated uncertainty for the measurement of the CP-violating 
$B^0_s$--$\bar B^0_s$ mixing phase. We point out that these effects 
can actually be controlled by means of an analysis of the time-dependent 
angular distribution of the $B^0_s\to J/\psi[\to\ell^+\ell^-]\bar K^{*0}[\to \pi^+K^-]$ 
decay products, and illustrate this through numerical studies. Moreover, 
we discuss $SU(3)$-breaking effects, which limit the theoretical accuracy of our 
method, and suggest internal consistency checks of $SU(3)$.
\end{abstract}

\vspace*{0.5truecm}
\vfill
\noindent
October 2008

\end{titlepage}

\thispagestyle{empty}
\vbox{}
\newpage

\setcounter{page}{1}

\section{Introduction}\label{sec:intro}
The exploration of CP-violating effects in $B_s$-meson decays offers a 
particularly promising probe for the search of New Physics (NP). In this
respect, a key channel is $B^0_s\to J/\psi\phi$, which is the counterpart
of the ``golden" decay $B^0_d\to J/\psi K_{\rm S}$ to measure the
angle $\beta$ in the unitarity triangle (UT) of the Cabibbo--Kobayashi--Maskawa
(CKM) matrix. Since the $B^0_s\to J/\psi\phi$ decay involves two vector
mesons in the final state, the time-dependent angular distribution 
of the decay products of the vector mesons, $J/\psi\to \ell^+\ell^-$ and 
$\phi\to K^+K^-$, has to be measured in order to disentangle the 
admixture of different CP eigenstates \cite{DDF,DFN}. 

Within the Standard Model (SM), the CP-violating effects in the time-dependent 
$B^0_s\to J/\psi\phi$ angular distribution are expected to be 
small. On the other hand, a preferred mechanism to accommodate a 
measurement of non-vanishing CP asymmetries would be given
by CP-violating NP contributions to $B^0_s$--$\bar B^0_s$ mixing
(see, for instance, \cite{BaFl}). Recent results from the first tagged, 
time-dependent $B^0_s\to J/\psi\phi$ analyses performed by the CDF 
\cite{CDF} and \D0 \cite{D0} collaborations at the Tevatron (FNAL) may 
actually point towards this direction, and have led 
to quite some attention \cite{UTfit}. The $B^0_s\to J/\psi\phi$ decay
is a main target of the LHCb experiment (CERN), which will soon start 
taking data and will allow us to explore the CP-violating phenomena in
this transition with impressive accuracy  \cite{LHCb}: already with
$2\,\mbox{fb}^{-1}$ of data, corresponding to one nominal year of
operation, the experimental uncertainty for the $B^0_s$--$\bar B^0_s$ mixing 
phase $\phi_s$ is expected to be $\sigma(\phi_s)_{\rm exp}\sim 1^\circ$, and 
an upgrade of LHCb with an integrated luminosity of $100\,\mbox{fb}^{-1}$ 
would eventually allow us to even reach a sensitivity of 
$\sigma(\phi_s)_{\rm exp}\sim 0.2^\circ$ \cite{LHCb-upgrade}.

In view of these exciting prospects, we have a closer look at the CP-violating 
effects in the time-dependent $B^0_s\to J/\psi[\to \ell^+\ell^-]\phi[\to K^+K^-]$ 
angular distribution that arise within the SM and limit the theoretical 
accuracy of the benchmark for the search for NP. Here the key r\^ole
is played by penguin topologies, which are doubly Cabibbo suppressed
and hence usually assumed to be negligible. However, these contributions 
cannot be calculated reliably from QCD, and could mimic CP-violating 
effects which might be misinterpreted as signals of NP in $B^0_s$--$\bar B^0_s$ 
mixing with a small but sizeable CP-violating NP phase.

In the present paper, we point out that the penguin effects can actually be 
controlled by means of an analysis of the angular distribution of 
$B^0_s\to J/\psi[\to \ell^+\ell^-]\bar K^{*0}[\to\pi^+ K^-]$ and its CP conjugate. 
Applying $SU(3)$ flavour-symmetry arguments and neglecting penguin 
annihilation and exchange topologies (which can be probed through 
$B^0_d\to J/\psi \phi$), the relevant hadronic parameters entering the 
$B^0_s\to J/\psi\phi$ observables can be determined, thereby allowing us 
to take them into account in the extraction of $\phi_s$.  We suggest to perform 
a simultaneous analysis of the $B^0_s\to J/\psi\phi$ and $B^0_s\to J/\psi \bar K^{*0}$ 
channels at LHCb, and encourage the CDF and \D0 collaborations to search for 
signals of this transition, as these would allow us to give first constraints on the 
penguin effects in $B^0_s\to J/\psi\phi$ and their impact on the extraction of the 
CP-violating $B^0_s$--$\bar B^0_s$ mixing phase. Further information can be 
obtained from the $B^0_d\to J/\psi\rho^0$ decay, in particular for the resolution 
of a discrete ambiguity through experimental data. 

As pointed out in Ref.~\cite{FFJM}, the data for CP violation in
$B^0_d\to J/\psi \pi^0$ and the branching ratio of this channel signal
that such effects are sizeable and soften the tension in the fit of
the UT between its angle $\beta$ and side $R_b$ as determined through 
CP violation in $B^0_d\to J/\psi K_{\rm S,L}$ decays and semileptonic 
$b\to u,c$ transitions, respectively. In particular, the measurement of $\beta$ 
has already reached a level of precision where subleading effects, i.e.\ doubly 
Cabibbo-suppressed penguin contributions, have to be included in order 
to match the experimental accuracy (see also Ref.~\cite{CPS}).
This feature strengthens the need to deal with such effects in analyses of CP violation 
in $B^0_s\to J/\psi \phi$ as well. In particular, we expect that the penguin effects
interfere constructively with mixing-induced CP violation and could lead
to CP asymmetries as large as ${\cal O}(-10\%)$, which would be significantly 
larger than the naive SM estimate of $\sin\phi_s^{\rm SM}\approx-3\%$ and
could be well detected at LHCb. 

The outline of this paper is as follows: in Section~\ref{sec:rev}, we give
an overview of the $B^0_s\to J/\psi[\to \ell^+\ell^-]\phi[\to K^+K^-]$ 
analysis and explore the impact of the penguin effects on the measurement
of $\phi_s$, while we discuss the strategy to include the hadronic penguin
contributions with the help of $B^0_s \to J/\psi \bar K^{*0}$ 
in Section~\ref{sec:contr}. This strategy is illustrated in Section~\ref{sec:illu}.
A detailed discussion of $SU(3)$-breaking effects and internal consistency checks 
that are offered by the observables of our decays into two vector mesons
are given in Section~\ref{sec:uncert}. Finally, we summarize our 
conclusions in Section~\ref{sec:concl}.

\boldmath
\section{Review of $B^0_s \to J/\psi \phi$}\label{sec:rev}
\unboldmath
\subsection{Structure of the Angular Distribution}
In contrast to the decay $B^0_d\to J/\psi K_{\rm S}$, we have
to deal with two vector mesons in the final state of $B^0_s \to J/\psi \phi$, 
which is  an admixture of CP-odd and CP-even eigenstates. 
Using the angular  distribution of the decay products of the vector mesons, 
the CP eigenstates can be disentangled. To this end, we introduce linear 
polarization states of the vector mesons, which are longitudinal ($0$) or 
transverse to their directions of motion. In the latter case, the polarization 
states may be parallel ($\parallel$) or perpendicular ($\perp$) to one another 
\cite{rosner}. The time-dependent angular distribution of 
$B^0_s\to J/\psi \phi$ takes the following general 
form \cite{DDF}:
\begin{equation}\label{ang-dist}
f(\Theta,\Phi,\Psi;t)=\sum_k \,
g^{(k)}(\Theta,\Phi,\Psi) \, b^{(k)}(t),
\end{equation}
where the decay kinematics is described by the $g^{(k)}(\Theta,\Phi,\Psi)$,
and the time-dependent coefficients $b^{(k)}(t)$ are given as 
\begin{equation}
\begin{array}{c}
\left|A_f(t)\right|^2 \quad (f\in\{0,\parallel,\perp\}),\\
\mbox{}\\
\mbox{Re}\{A_0^\ast(t)A_\parallel(t)\}, \quad
\mbox{Im}\{A_f^\ast(t)A_\perp(t)\} \quad (f\in\{0,\parallel\}),
\end{array}
\end{equation}
with linear polarization amplitudes 
$A_f=\langle(J/\psi\phi)_{f}|{\cal H}_{\rm eff}|B^0_s(t)\rangle$, where
${\cal H}_{\rm eff}$ is the relevant low-energy effective Hamiltonian.
Here $A_\perp(t)$ describes a CP-odd final-state configuration, whereas
$A_0(t)$ and $A_\parallel(t)$ correspond to CP-even final-state configurations.

In the case of the CP-conjugate decay $\bar B^0_s\to J/\psi \phi$, we may
write the angular distribution as
\begin{equation}\label{ang-CP}
\bar f(\Theta,\Phi,\Psi;t)=\sum_k\bar {\cal O}^{(k)}(t)
g^{(k)}(\Theta,\Phi,\Psi).
\end{equation}
Since the meson content of the $ J/\psi \phi$ state is the 
same whether it results from the $B_s^0$ or $\bar B_s^0$ decays, 
we may use the same angles $\Theta$, $\Phi$ and $\Psi$ 
as in (\ref{ang-dist}) to describe the kinematics of the decay products. 
Following these lines, the effects of CP transformations relating 
$B_s^0\to (J/\psi \phi)_f$ to $\bar B_s^0\to (J/\psi \phi)_f$ are then
taken into through the CP eigenvalues of the final-state
configuration $(J/\psi \phi)_f$. Therefore the same functions 
$g^{(k)}(\Theta,\Phi,\Psi)$ are present in (\ref{ang-dist}) and (\ref{ang-CP}). 
For the explicit form the of these quantities, see Ref.~\cite{DDF}.

\begin{figure}
   \centering
   \includegraphics[width=5.5truecm]{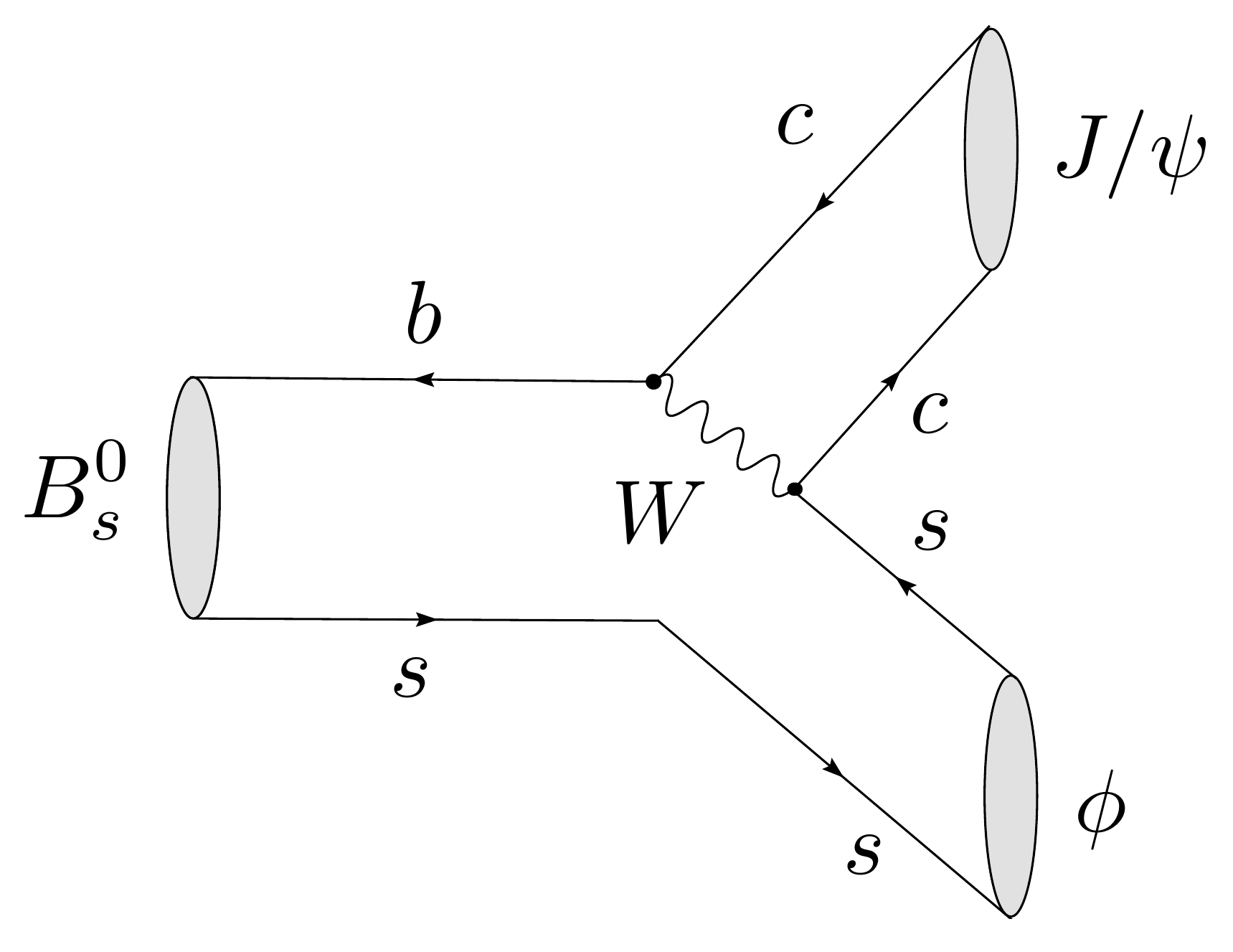}
   \hspace*{0.5truecm} 
   \includegraphics[width=5.5truecm]{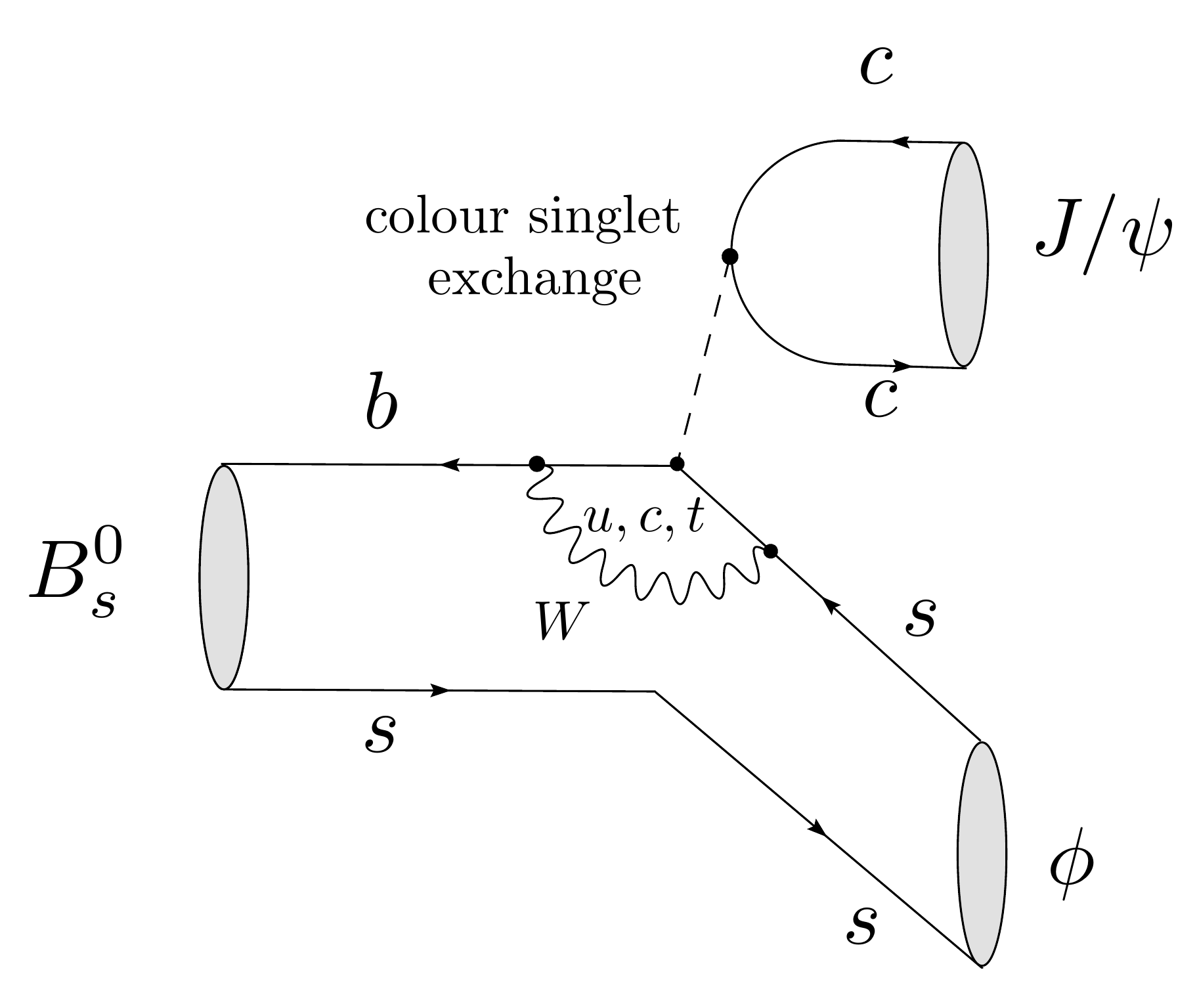} 
   \caption{Decay topologies contributing to $B^0_s \to J/\psi \phi$ in the SM.}\label{fig:1}
\end{figure}

\subsection{Structure of the Decay Amplitudes}
As can be seen in Fig.~\ref{fig:1}, colour-suppressed tree-diagram-like
and penguin topologies contribute to the $B^0_s \to J/\psi \phi$ decay
within the SM. For a given final-state configuration $f\in\{0,\parallel,\perp\}$, 
the $B_s^0\to J/\psi \phi$ decay amplitude can therefore be written as
\begin{equation}\label{Bspsiphi-ampl}
A(B_s^0\to (J/\psi \phi)_f)=
{\lambda_c^{(s)}}{\left[A_{{\rm T}}^{(c)f}+
A_{{\rm P}}^{(c)f}\right]}+{\lambda_u^{(s)}}
{A_{{\rm P}}^{(u)f}}
+{\lambda_t^{(s)}}{A_{{\rm P}}^{(t)f}},
\end{equation}
where the $\lambda_j^{(s)}\equiv  V_{js} V_{jb}^\ast$ are
CKM factors, while $A_{{\rm T}}^{(c)f}$ and $A_{{\rm P}}^{(j)f}$ are 
CP-conserving strong amplitudes related to tree-diagram-like and penguin 
topologies (with internal $j\in\{u,c,t\}$ quarks), respectively. Using the
appropriate low-energy effective Hamiltonian, the latter quantities can be 
expressed in terms of linear combinations of perturbatively calculable 
Wilson coefficient functions and non-perturbative hadronic matrix elements 
of the corresponding four-quark operators, which are associated with large
uncertainties. Using the CKM unitarity relation 
$\lambda_t^{(s)}=-\lambda_c^{(s)}-\lambda_u^{(s)}$ to eliminate the 
$\lambda_t^{(s)}$ factor, we obtain
\begin{equation}\label{ampl}
A(B_s^0\to (J/\psi \phi)_f)=\left(1-\frac{\lambda^2}{2}\right)
{\cal A}_f\left[1+\epsilon a_f e^{i\theta_f}e^{i\gamma}\right],
\end{equation}
where 
\begin{equation}\label{Amplf-def}
{\cal A}_f\equiv \lambda^2 A \left[A_{{\rm T}}^{(c)f}+A_{{\rm P}}^{(c)f}-
A_{{\rm P}}^{(t)f}\right]
\end{equation}
and
\begin{equation}\label{af-def}
a_f e^{i\theta_f}\equiv R_b
\left[\frac{A_{{\rm P}}^{(u)f}-A_{{\rm P}}^{(t)f}}{A_{{\rm T}}^{(c)f}+
A_{{\rm P}}^{(c)f}-A_{{\rm P}}^{(t)f}}\right]
\end{equation}
are CP-conserving hadronic parameters, while
\begin{eqnarray}
\lambda\equiv|V_{us}|&=&0.22521\pm0.00083, \\
A \equiv  |V_{cb}|/\lambda^2&=&0.809\pm0.026, \\
R_b  \equiv  (1-\lambda^2/2)|V_{ub}/(\lambda V_{cb})|&=&
0.423 ^{+0.015}_{-0.022}\pm 0.029,\\
\epsilon \equiv \lambda^2/(1-\lambda^2)&=&0.053
\end{eqnarray}
are CKM parameters \cite{FFJM,PDG}, and the UT angle $\gamma$ flips its sign 
when considering CP-conjugate processes:
\begin{equation}\label{ampl-CP}
A(\bar B_s^0\to (J/\psi \phi)_f)=\eta_f\left(1-\frac{\lambda^2}{2}\right)
{\cal A}_f\left[1+\epsilon a_f e^{i\theta_f}e^{-i\gamma}\right].
\end{equation}
Here $\eta_f$ is the CP eigenvalue of the final-state configuration $(J/\psi \phi)_f$.

\subsection{Time-dependent Observables}
If we neglect CP violation in $B^0_s$--$\bar B^0_s$ oscillations, which can
be probed through wrong-charge lepton asymmetries and is a tiny effect 
in the SM, the formalism of $B^0_s$--$\bar B^0_s$ mixing yields the following
expressions \cite{RF-ang}:
\begin{equation}\label{untagged}
\Gamma[f,t]
\equiv|A_f(t)|^2+|\overline{A}_f(t)|^2=R_{\rm L}^f\,e^{-\Gamma_{\rm L}^{(s)}t}+
R_{\rm H}^f\,e^{-\Gamma_{\rm H}^{(s)}t} ,
\end{equation}
\begin{equation}\label{tagged}
|A_f(t)|^2-|\overline{A}_f(t)|^2=2\,e^{-\Gamma_s t}\left[
A_{\rm D}^f\cos(\Delta M_st) + A_{\rm M}^f \sin(\Delta M_st)\right],
\end{equation}
where $\Gamma_{\rm L}^{(s)}$ and $\Gamma_{\rm H}^{(s)}$ are
the decay widths of the ``light" and ``heavy" $B_s$ mass eigenstates, 
respectively, $\Gamma_s$ is their average, and 
$\Delta M_s\equiv M_{\rm H}^{(s)}-M_{\rm L}^{(s)}$ the difference of the
mass eigenvalues. The labels ``D" and ``M" remind us that non-vanishing values of
$A_{\rm D}^f$ and $A_{\rm M}^f$ are generated through direct and mixing-induced
CP-violating effects, respectively. 

Since the hadronic parameters $a_f e^{i\theta_f}$, which are essentially
unknown, enter (\ref{ampl}) and (\ref{ampl-CP}) in combination with the
doubly Cabibbo-suppressed parameter $\epsilon$, they are usually
neglected. In this limit, we obtain 
\begin{equation}\label{rate-untagged}
\Gamma[f,t]=|{\cal N}_f|^2\left[ 
(1+\eta_f\cos\phi_s)e^{-\Gamma_{\rm L}^{(s)}t}+
(1-\eta_f\cos\phi_s)e^{-\Gamma_{\rm H}^{(s)}t}\right],
\end{equation}
\begin{equation}\label{rate-diff}
|A_f(t)|^2-|\overline{A}_f(t)|^2=2\eta_f |{\cal N}_f|^2
e^{-\Gamma_s t}\sin\phi_s \sin(\Delta M_st),
\end{equation}
where we have introduced the abbreviation ${\cal N}_f\equiv(1-\lambda^2/2){\cal A}_f'$, 
and $\phi_s$ is the CP-violating $B^0_s$--$\bar B^0_s$ mixing phase. In the 
ratio of the CP-violating rate difference (\ref{rate-diff}), which requires the
``tagging" of whether we had an initially, i.e.\ at time $t=0$, 
present $B^0_s$ or $\bar B^0_s$ meson, and the ``untagged" rate 
(\ref{rate-untagged}), the overall normalization $|{\cal N}_f|$ cancels, so that $\phi_s$ 
can be extracted. For the corresponding time-dependences of the other observables 
provided by the angular distribution, see Ref.~\cite{DFN}. 

In the SM, we have $\phi_s^{\rm SM}=-2\lambda^2\eta=-(2.12\pm0.11)^\circ$,
where the numerical value follows from the current CKM fits \cite{CKMfits}.
However, since $B^0_s$--$\bar B^0_s$ mixing is a strongly suppressed
flavour-changing neutral-current (FCNC) process in the SM, it is a sensitive
probe for NP effects in the TeV regime. Should new particles actually contribute
to this phenomenon, the off-diagonal mass element of the mixing matrix is 
modified as follows \cite{BaFl}:
\begin{equation}
M_{12}^s = M_{12}^{s,{\rm SM}} \left(1 + \kappa_s e^{i\sigma_s}\right),
\end{equation}
where $\kappa_s$ measures the strength of the NP contribution with respect
to the SM, and $\sigma_s$ is a CP-violating NP phase. Consequently, we have
\begin{equation}
\Delta M_s   =  \Delta M_s^{\rm SM}\left| 1 + \kappa_s  e^{i\sigma_s}\right|,
\end{equation}
\begin{equation}
\phi_s  =   \phi_s^{\rm SM}+\phi_s^{\rm NP}=
-2\lambda^2\eta + \arg (1+\kappa_s e^{i\sigma_s}).
\end{equation}
As discussed in Ref.~\cite{BaFl}, the values of 
$\rho_s\equiv \Delta M_s/\Delta M_s^{\rm SM}$ and $\phi_s^{\rm NP}$
can be converted into contours in the $\sigma_s$--$\kappa_s$ plane, 
which sets the parameter space for NP contributions to $B^0_s$--$\bar B^0_s$ mixing.

For many years, only lower bounds on $\Delta M_s$ 
were available from the LEP (CERN) experiments and SLD (SLAC) \cite{LEPBOSC}. 
In 2006, the value of $\Delta M_s$ could eventually be pinned down at the Tevatron 
\cite{DMs-obs}. The current status can be summarized as follows:
\begin{equation}\label{MDs}
\Delta M_s=\left\{
\begin{array}{ll}
(18.56 \pm0.87){\rm ps}^{-1}  & \mbox{(\D0 collaboration \cite{D0-DMs}),} \\
(17.77\pm0.10 \pm 0.07){\rm ps}^{-1} & 
\mbox{(CDF collaboration \cite{CDF-DMs})}.
\end{array}
\right.
\end{equation}
In order to determine the parameter $\rho_s$ from these measurements, 
the SM value of $\Delta M_s$ is required, involving a hadronic 
parameter $f_{B_s}^2\hat B_{B_s}$, which can be determined by means of 
lattice QCD techniques and introduces the corresponding uncertainties into
the analysis. The HPQCD collaboration finds 
$\Delta M_s^{\rm SM}=20.3(3.0)(0.8)\,{\rm ps}^{-1}$ \cite{HPQCD-DMs}, which 
yields $\rho_s=0.88\pm0.13$.

\begin{figure}
   \centering
   \includegraphics[width=7.0truecm]{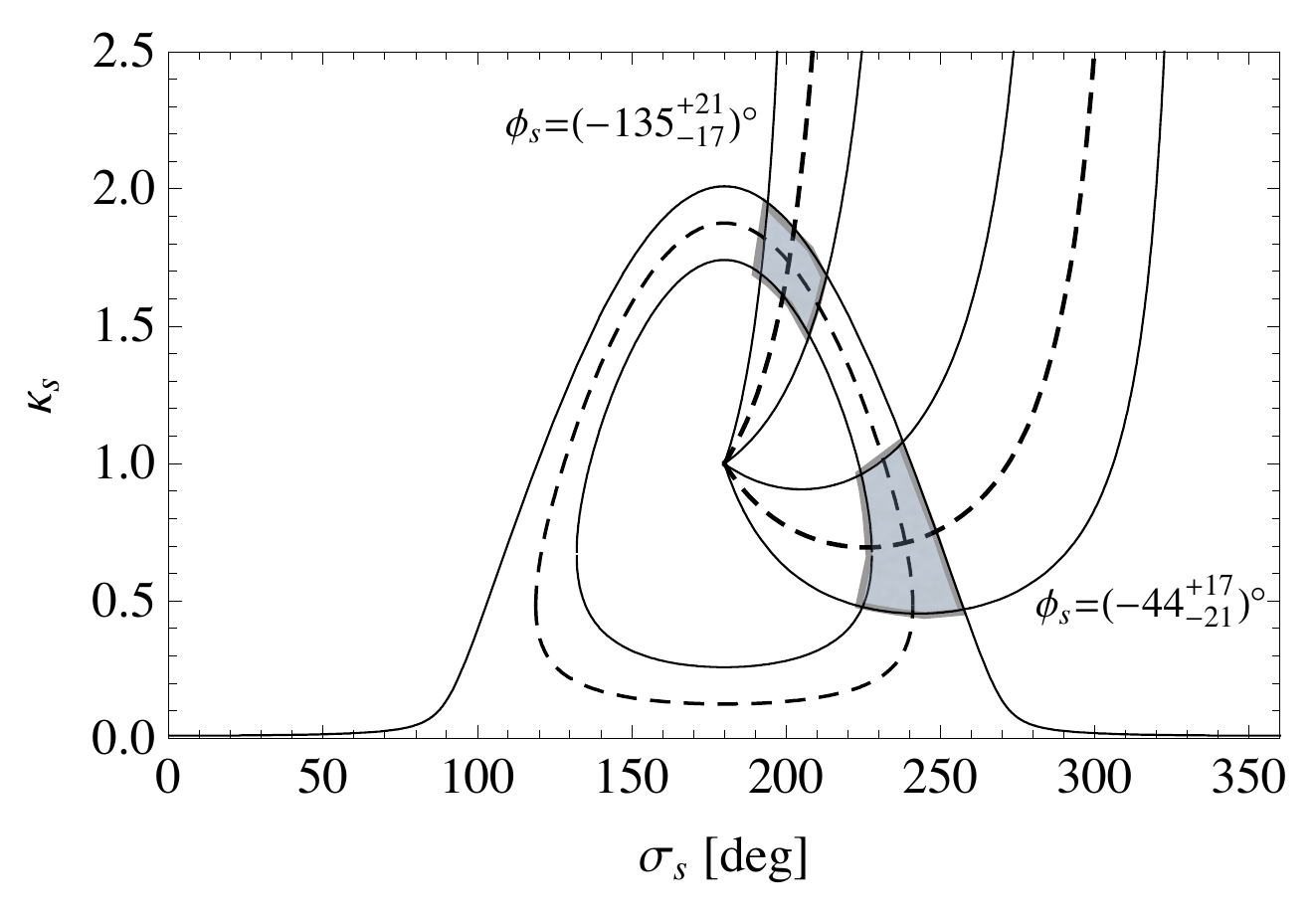} 
   \caption{The situation in the $\sigma_s$--$\kappa_s$ plane of the
   NP parameters for $B^0_s$--$\bar B^0_s$ mixing arising from current data
   and theory input, as discussed in the text.}\label{fig:2}
\end{figure}

Recently, following Refs.~\cite{DDF,DFN}, the CDF  \cite{CDF} and \D0 \cite{D0} collaborations have reported the first results from tagged, time-dependent analyses
of the full three-angle distribution of the $B^0_s\to J/\psi[\to \ell^+\ell^-]\phi[\to K^+K^-]$
decay products. In an analysis by the UTfit collaboration \cite{UTfit}, taking also other 
constraints into account, it is argued that these results may indicate CP-violating 
NP contributions to $B^0_s$--$\bar B^0_s$ mixing, which would immediately rule 
out models with minimal flavour violation (MFV). Recently, a first average of 
the CDF and \D0 data was presented by the Heavy Flavour Averaging Group
(HFAG) \cite{HFAG}, corresponding to the following twofold solution:
\begin{equation}\label{phis-UTfit}
\phi_s=\left(-44^{+17}_{-21}\right)^\circ \, \lor \,
\left(-135^{+21}_{-17}\right)^\circ.
\end{equation}
In Fig.~\ref{fig:2}, we show -- as an update of the analysis performed in 
Ref.~\cite{BaFl} -- the corresponding situation in the $\sigma_s$--$\kappa_s$ 
plane: the central hill-like region corresponds to $\rho_s$, i.e.\ the mass 
difference $\Delta M_s$, while the two branches represent the twofold solution 
for $\phi_s$; the overlap of the $\Delta M_s$ and $\phi_s$ constraints results
in the two shaded allowed regions. It will be very interesting to monitor these 
measurements in the future. Fortunately, the  $B^0_s\to J/\psi\phi$ analyses 
are very accessible at the LHCb experiment \cite{LHCb}, which will soon start 
taking data.

\subsection{Impact of Penguin Contributions}\label{ssec:phis-impact}
The experimental results discussed in the previous section were obtained
by assuming that the doubly Cabibbo-suppressed parameters
$a_f e^{i\theta_f}$, which describe -- sloppily speaking -- the ratio of
penguin to tree contributions, play a negligible r\^ole. In view of the
search for NP signals, which requires a solid control of the SM effects, 
and the tremendous accuracy that can be achieved at LHCb, we 
generalize here the formulae to take also these contributions into account. 

Let us first have a look at the untagged observables. Following 
Ref.~\cite{RF-ang}, we have
\begin{eqnarray}
R_{\rm L}^f&=&|{\cal N}_f|^2\Bigl[(1+\eta_f\cos\phi_s)\label{RL-gen}\\
&&\hspace*{-0.9truecm}+2\epsilon a_f\cos\theta_f\left\{\cos\gamma+
\eta_f\cos(\phi_s+\gamma)\right\}
+\epsilon^2a_f^2\left\{1+\eta_f\cos(\phi_s+2\gamma)\right\}\Bigr],\nonumber
\end{eqnarray}
\begin{eqnarray}
R_{\rm H}^f&=&|{\cal N}_f|^2\Bigl[(1-\eta_f\cos\phi_s)\label{RH-gen}\\
&&\hspace*{-0.9truecm}+2 \epsilon a_f\cos\theta_f\left\{\cos\gamma-
\eta_f\cos(\phi_s+\gamma)\right\}+
\epsilon^2a_f^2\left\{1-\eta_f\cos(\phi_s+2\gamma)\right\}\Bigr],\nonumber
\end{eqnarray}
so that
\begin{equation}\label{untag-0}
\Gamma[f,t=0]=R_{\rm L}^f+R_{\rm H}^f=2|{\cal N}_f|^2\Bigl[1+
2\epsilon a_f\cos\theta_f\cos\gamma+\epsilon^2a_f^2\Bigr].
\end{equation}
On the other hand, the CP-violating observables are given as follows:
\begin{eqnarray}
A_{\rm D}^f&\hspace*{-0.2truecm}=\hspace*{-0.2truecm}&
-2 |{\cal N}_f|^2 \epsilon a_f\sin\theta_f \sin\gamma,
\label{AD-expr}\\
A_{\rm M}^f&\hspace*{-0.2truecm}=\hspace*{-0.2truecm}&\eta_f |{\cal N}_f|^2
\left[\sin\phi_s+2\epsilon a_f\cos\theta_f\sin(\phi_s+\gamma)+\epsilon^2
a_f^2\sin(\phi_s+2\gamma)\right].\label{AM-expr}
\end{eqnarray}
Note that (\ref{RL-gen}) and (\ref{RH-gen}) are not independent from
(\ref{AD-expr}) and (\ref{AM-expr}), as
\begin{equation}
\bigl(A_{\rm D}^f\bigr)^2+\bigl(A_{\rm M}^f\bigr)^2=R_{\rm L}^fR_{\rm H}^f\,.
\end{equation}

\begin{figure}
   \centering
   \begin{tabular}{cc}
   \includegraphics[width=6.5truecm]{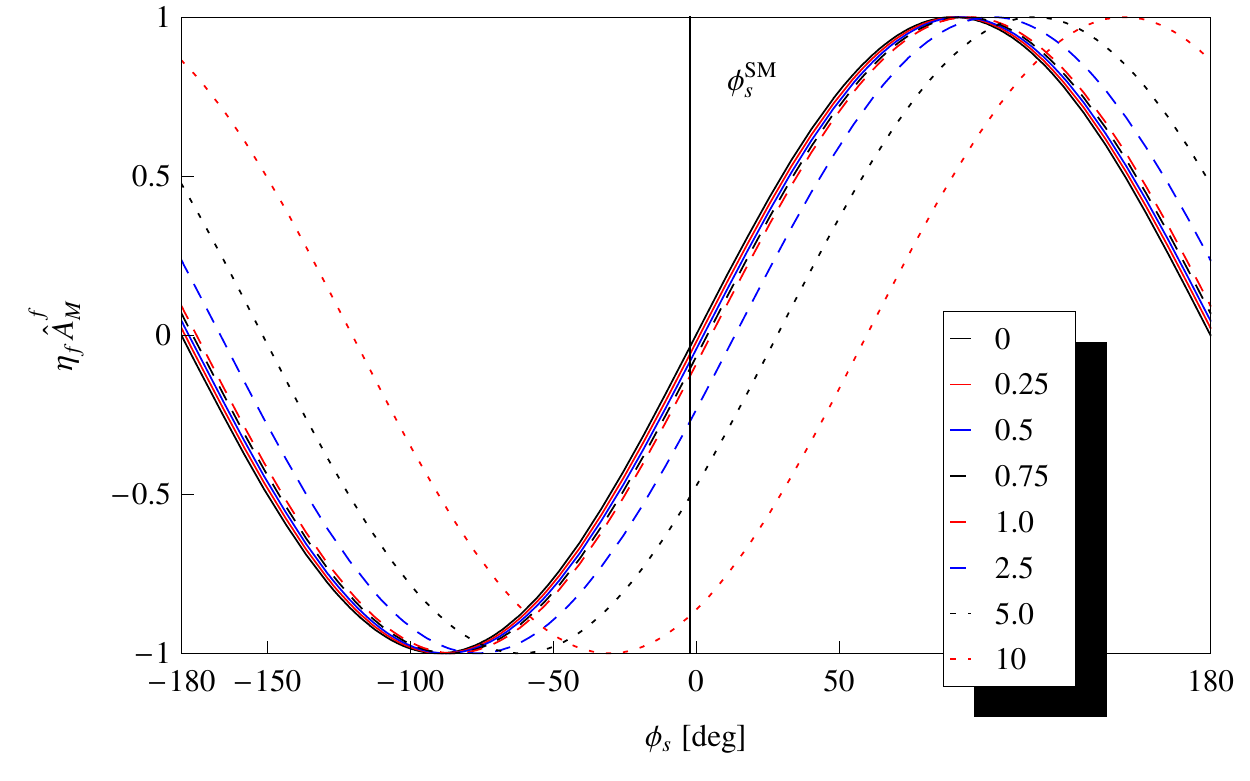} &
    \includegraphics[width=6.5truecm]{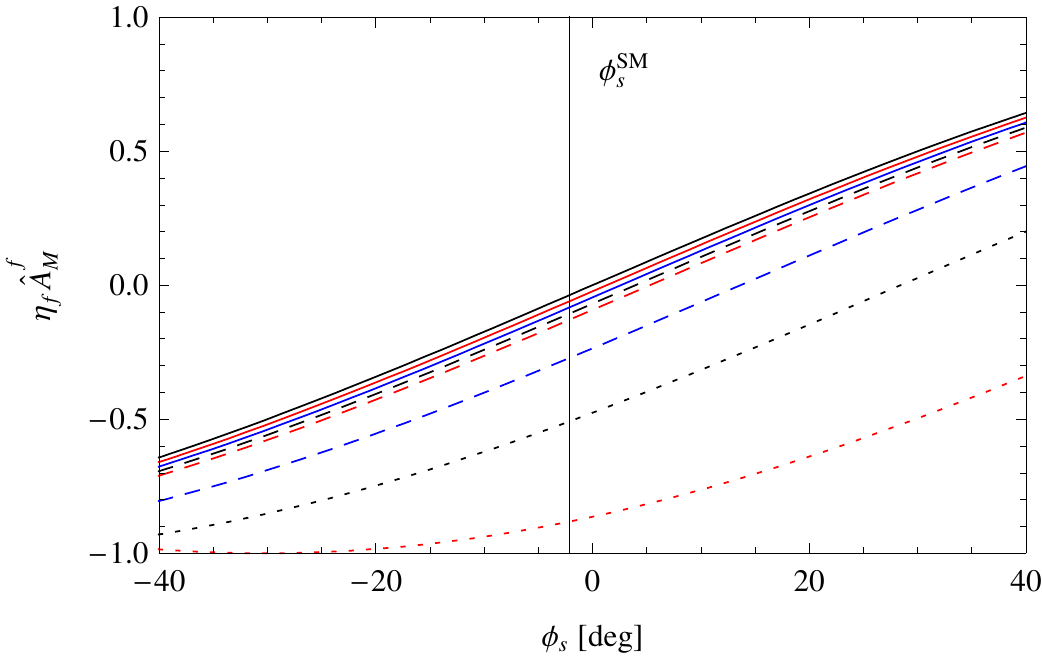} 
    \end{tabular}
   \caption{Impact of the penguin parameter $a_f$  on the 
   mixing-induced CP asymmetry $\eta_f\hat A_{\rm M}^f$ for $\theta_f=180^\circ$
  as function of the $B^0_s$--$\bar B^0_s$ mixing phase $\phi_s$.}\label{fig:3}
\end{figure}

The ratio of the ``tagged" rate difference (\ref{tagged}) and the ``untagged" rate  
(\ref{untagged}) can be written as
\begin{equation}\label{t-dep-asym}
\frac{|A_f(t)|^2-|\overline{A}_f(t)|^2}{|A_f(t)|^2+|\overline{A}_f(t)|^2}=
\frac{\hat A_{\rm D}^f\cos(\Delta M_st) + \hat A_{\rm M}^f 
\sin(\Delta M_st)}{\cosh(\Delta\Gamma_st/2)-{\cal A}_{\Delta\Gamma}^f
\sinh(\Delta\Gamma_st/2)} ,
\end{equation}
where $\Delta\Gamma_s\equiv\Gamma_{\rm H}^{(s)}-\Gamma_{\rm L}^{(s)}$, and 
\begin{equation}
{\cal A}_{\Delta\Gamma}^f=\frac{R_{\rm H}^f-R_{\rm L}^f}{R_{\rm H}^f+R_{\rm L}^f}.
\end{equation}
If we introduce
\begin{equation}
N_f\equiv1+2\epsilon a_f\cos\theta_f\cos\gamma+\epsilon^2a_f^2
=\frac{\Gamma[f,t=0]}{2|{\cal N}_f|^2},
\end{equation}
the corresponding observables take the following forms:
\begin{equation}\label{AD-form}
\hat A_{\rm D}^f=\frac{-2 \epsilon a_f\sin\theta_f \sin\gamma}{N_f},
\end{equation}
\begin{equation}\label{AM-form}
 \hat A_{\rm M}^f =+\frac{\eta_f}{N_f}\left[\sin\phi_s+2\epsilon a_f\cos\theta_f
 \sin(\phi_s+\gamma)+\epsilon^2 a_f^2\sin(\phi_s+2\gamma)\right],
\end{equation}
\begin{equation}\label{ADG-form}
{\cal A}_{\Delta\Gamma}^f=-\frac{\eta_f}{N_f}\left[\cos\phi_s+2\epsilon 
 a_f\cos\theta_f\cos(\phi_s+\gamma)+\epsilon^2 a_f^2
 \cos(\phi_s+2\gamma)\right].
\end{equation}
The measurement of ${\cal A}_{\Delta\Gamma}^f$ relies on a sizeable value
of the width difference $\Delta\Gamma_s$. Moreover, we have
\begin{equation}
\bigl(\hat A_{\rm D}^f\bigr)^2+\bigl(\hat A_{\rm M}^f\bigr)^2+
\bigl({\cal A}_{\Delta\Gamma}^f\bigr)^2=1.
\end{equation}
For the extraction of $\phi_s$, the key observables are the $ \hat A_{\rm M}^f$;
in Fig.~\ref{fig:3}, we illustrate the impact of the penguin parameter $a_f$.
Since $a_fe^{i\theta_f}$ is defined in (\ref{af-def}) in such a way that
$\theta_f$ is given by $180^\circ$ if we assume factorization, we have
used this value in order to calculate the curves shown in Fig.~\ref{fig:3}. For this
strong phase the penguin effects are actually maximal in $ \hat A_{\rm M}^f$
since only $\cos\theta_f$ enters. On the other hand, the direct CP asymmetries
$\hat A_{\rm D}^f$ would then vanish, as they are proportional to $\sin\theta_f$.

We observe that in order to accommodate a value of 
$\phi_s\sim-44^\circ$, as given in (\ref{phis-UTfit}), we 
would need $a_f\sim2.5\mbox{--}5$, which appears completely unrealistic. 
However, since $a_f$ suffers from large uncertainties, values as large as 
$0.5\sim1$ can a priori not be excluded. Should $\phi_s$ take a value on 
the small side, these hadronic SM contributions would lead to a significant 
uncertainty in the extraction of the $B^0_s$--$\bar B^0_s$ mixing phase.

\begin{figure}
   \centering
   \begin{tabular}{cc}
   \includegraphics[width=6.5truecm]{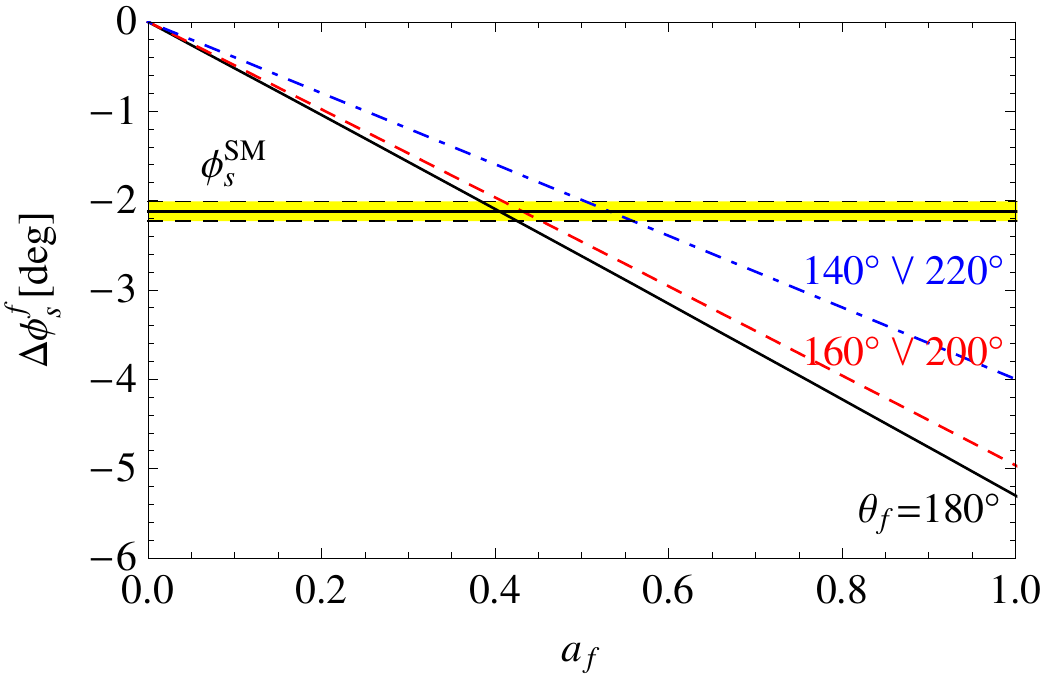} &
    \includegraphics[width=6.5truecm]{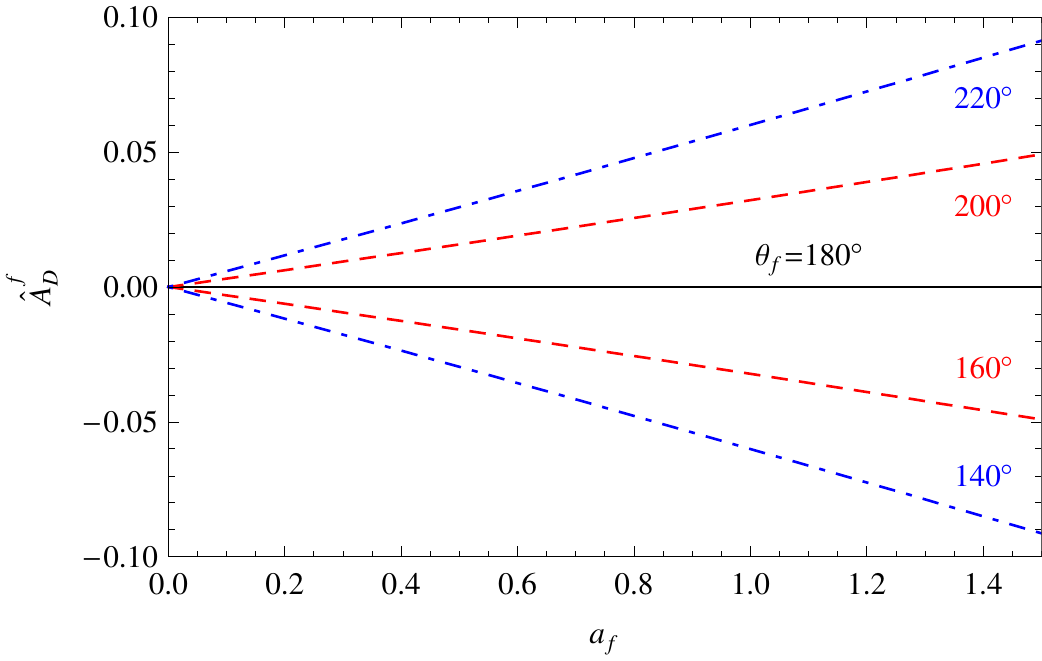} 
    \end{tabular}
   \caption{Left panel: the dependence of $\Delta\phi^f_s$ on $a_f$ for various 
   values of $\theta_f$; right panel: the dependence of $\hat A_{\rm D}^f$ on 
   $a_f$ for various values of $\theta_f$.}\label{fig:4}
\end{figure}

In order to explore this effect in more detail, we use (\ref{AD-form}) and 
(\ref{AM-form}) to derive the following expression:
\begin{equation}\label{AM-pen-expr}
\frac{\eta_f \hat A_{\rm M}^f}{\sqrt{1-(\hat A_{\rm D}^f)^2}}=\sin(\phi_s+\Delta\phi^f_s),
\end{equation}
where
\begin{equation}\label{sDelphis}
\sin\Delta\phi^f_s=\frac{2 \epsilon a_f\cos\theta_f \sin\gamma+\epsilon^2a_f^2
\sin2\gamma}{N_f\sqrt{1-(\hat A_{\rm D}^f)^2}}
\end{equation}
and
\begin{equation}\label{cDelphis}
\cos\Delta\phi^f_s=\frac{1+ 2 \epsilon a_f\cos\theta_f \cos\gamma+\epsilon^2a_f^2
\cos2\gamma}{N_f\sqrt{1-(\hat A_{\rm D}^f)^2}},
\end{equation}
so that
\begin{equation}\label{tDelphis}
\tan\Delta\phi^f_s=\frac{2 \epsilon a_f\cos\theta_f \sin\gamma+\epsilon^2a_f^2
\sin2\gamma}{1+ 2 \epsilon a_f\cos\theta_f\cos\gamma+\epsilon^2a_f^2\cos2\gamma}.
\end{equation}
It should be stressed that the shift $\Delta\phi_s^f$ of the $B^0_s$--$\bar B^0_s$ 
mixing phase does not depend on the value of $\phi_s$ itself. 
In Fig.~\ref{fig:4}, we show the dependence of $\Delta\phi^f_s$ on the 
penguin parameter $a_f$ for various values of $\theta_f$, and give -- in 
order to monitor the corresponding direct CP asymmetries -- a similar plot 
for $\hat A_{\rm D}^f$. We observe that $\Delta\phi^f_s$ is of the same
size as $\phi_s^{\rm SM}$ for $a_f\sim0.4$, and that a value of $a_f\sim1$
would induce a shift of $\Delta\phi^f_s\sim-5^\circ$. As can be seen in the left 
panel of Fig.~\ref{fig:4}, we have $-0.05 \lsim \hat A_{\rm D}^f \lsim +0.05 $ for 
$a_f\lsim1$ and values of $|\theta_f-180^\circ|$ as large as $40^\circ$.
Interestingly, as we expect $\cos\theta_f<0$, the shift of $\phi_s$ is expected
to be negative as well, i.e.\ it would interfere constructively with $\phi_s^{\rm SM}$.
These features are fully supported by our recent analysis of the $B^0\to J/\psi \pi^0$
channel \cite{FFJM}. Consequently, it is important to get a handle on the 
penguin effects in the $B^0_s\to J/\psi\phi$ decay.

\boldmath
\section{The Control Channel $B^0_s \to J/\psi \bar K^{*0}$}\label{sec:contr}
\unboldmath
\subsection{Structure of the Decay Amplitudes}
\begin{figure}
   \centering
   \includegraphics[width=5.5truecm]{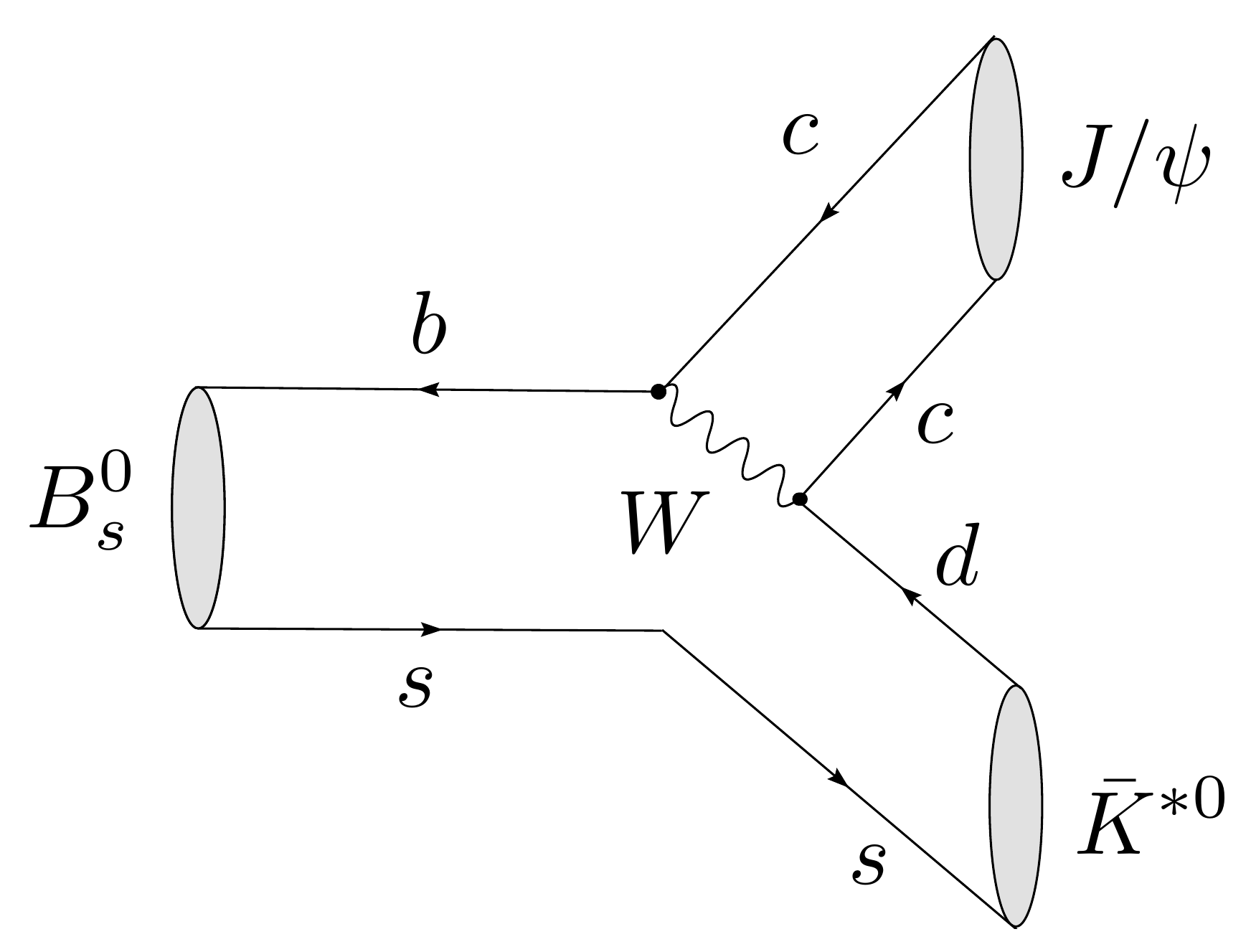}
   \hspace*{0.5truecm} 
   \includegraphics[width=5.5truecm]{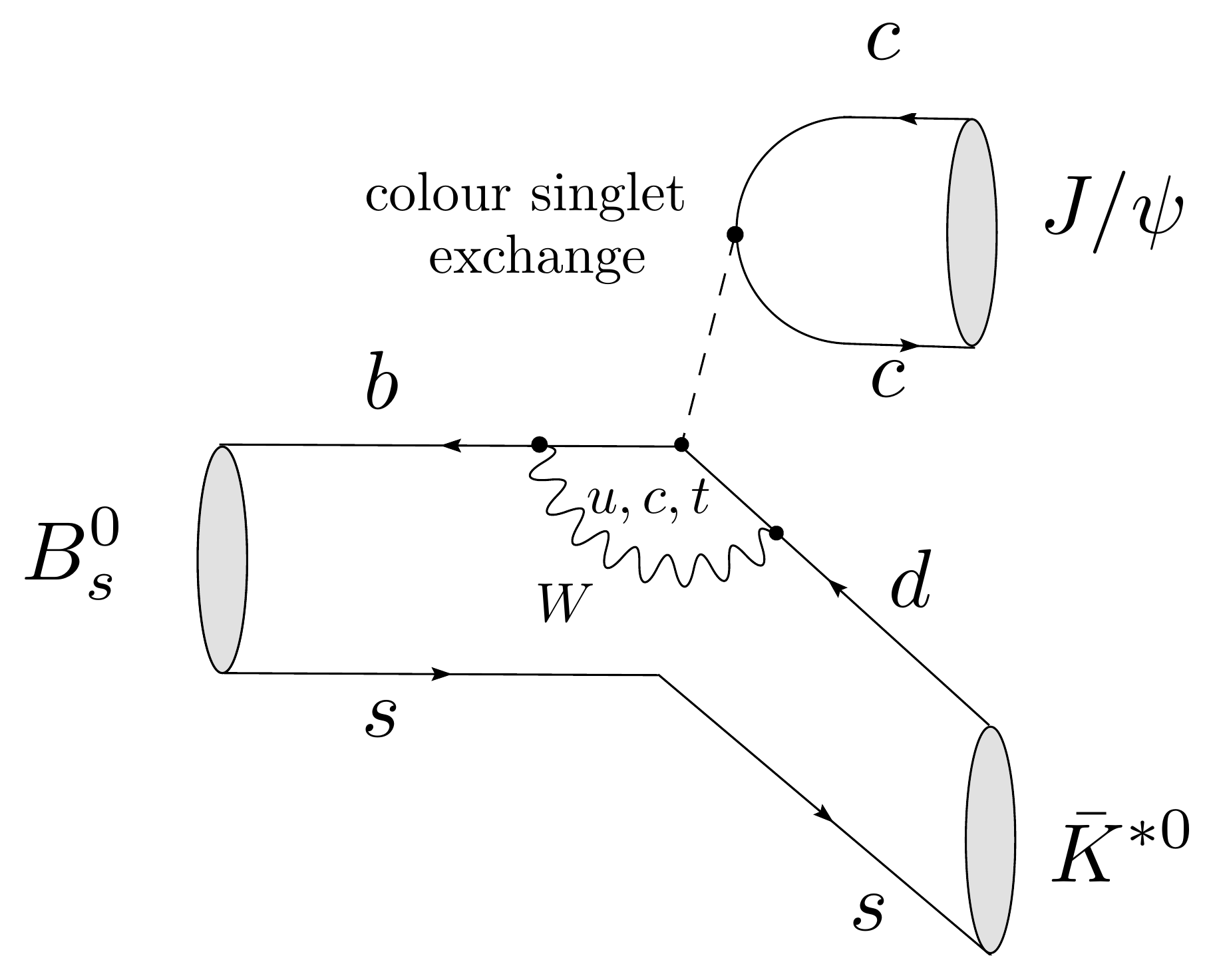} 
   \caption{Decay topologies contributing to $B^0_s \to J/\psi \bar K^{*0}$ 
   in the SM.}\label{fig:5}
\end{figure}

In Fig.~\ref{fig:5}, we show the decay topologies contributing to the
$B^0_s \to J/\psi \bar K^{*0}$ channel. The key difference with respect
to the $B^0_s \to J/\psi \phi$ decay shown in Fig.~\ref{fig:1} is that
$B^0_s \to J/\psi \bar K^{*0}$ is caused by $\bar b\to\bar d c\bar c$ quark-level
processes, whereas $B^0_s \to J/\psi \phi$ originates from  $\bar b\to\bar s c\bar c$ 
transitions. Consequently, the CKM factors are different in these channels.
In analogy to (\ref{ampl}), we may write
\begin{equation}\label{ampl-d}
A(B_s^0\to (J/\psi \bar K^{*0})_f)=\lambda{\cal A}_f'\left[1-a_f' e^{i\theta_f'}
e^{i\gamma}\right],
\end{equation}
where ${\cal A}_f'$ and $a_f' e^{i\theta_f'}$ are the counterparts of the
$B_s^0\to (J/\psi \phi)_f$ parameters introduced in (\ref{Amplf-def}) and 
(\ref{af-def}), respectively. In contrast to (\ref{ampl}), the latter parameter
does {\it not} enter (\ref{ampl-d}) in a doubly Cabibbo-suppressed way.
Consequently, the $B^0_s \to J/\psi \bar K^{*0}$ channel offers a sensitive 
probe for this quantity. If we apply the $SU(3)$ flavour symmetry of strong
interactions, we obtain
\begin{equation}\label{SU3-1}
|{\cal A}_f| = |{\cal A}_f'|,
\end{equation}
as well as
\begin{equation}\label{SU3-2}
a_f = a_f', \quad \theta_f = \theta_f'.
\end{equation}
In addition to $SU(3)$ flavour-symmetry arguments we have here also 
assumed that penguin annihilation ($PA$) and exchange ($E$) topologies, 
which contribute to $B_s^0\to (J/\psi \phi)_f$ but have no counterpart in 
$B^0_s \to J/\psi \bar K^{*0}$, play a negligible r\^ole. Fortunately, the 
importance of these topologies can be probed with the help of the 
$B^0_d\to (J/\psi \phi)_f$  channel, which has amplitudes proportional 
to $(PA+E)_f$. The Belle collaboration has recently reported the new
upper bound of $\mbox{BR}(B^0_d \to J/\psi \phi) < 9.4 \times 10^{-7}$ ($90\%$ C.L.)
\cite{Belle-psiphi}, which does not show any anomalous enhancement. 
The theoretical uncertainties associated with the application of the 
$SU(3)$ flavour symmetry will be discussed separately in 
Section~\ref{sec:uncert}.

\subsection{Observables}
In contrast to the $B^0_s \to J/\psi  [\to\ell^+\ell^-] \phi[\to K^+ K^-]$ decay, 
the final states of $B^0_s \to J/\psi [\to\ell^+\ell^-] \bar K^{*0}[\to\pi^+ K^-]$  
and its CP conjugate are flavour-specific, i.e.\ the charges of the pions and 
kaons depend on whether we had a
$B^0_s$ or $\bar B^0_s$ meson in the initial state. Consequently, 
the time-dependent angular distributions do not show 
CP violation due to interference between mixing and decay, i.e.\ the 
$A_{\rm M}^f$ observables introduced in (\ref{tagged}) have no counterparts,
and do not depend on the $B^0_s$--$\bar B^0_s$ mixing phase. However, 
untagged observables, as well as direct CP-violating asymmetries 
provide actually sufficient information to determine $a_f'$ and $\theta_f'$. 
In Appendix~\ref{app-a}, we give the expressions for the time-dependent 
angular distributions, which allow the determination of the relevant observables.

Let us first discuss the untagged case, and introduce 
\begin{equation}\label{H-def}
H_f\equiv \frac{1}{\epsilon}\left|\frac{{\cal A}_f}{{\cal A}'_f}\right|^2
\frac{\Gamma[f,t=0]'}{\Gamma[f,t=0]}=\frac{1-2a_f'\cos\theta_f'\cos\gamma
+a_f'^2}{1+2\epsilon a_f\cos\theta_f\cos\gamma+\epsilon^2a_f^2},
\end{equation}
where $\Gamma[f,t=0]'$ is the $B^0_s \to J/\psi \bar K^{*0}$ counterpart
of (\ref{untag-0}). Using (\ref{SU3-1}), we may extract $H_f$ from the 
untagged observables. Moreover, using also (\ref{SU3-2}), we can
determine $a_f'$ as a function of $\theta_f'$ with the help of the
following formulae:
\begin{equation}\label{af-Hdet}
a_f'=U_{H_f}\pm\sqrt{U_{H_f}^2-V_{H_f}},
\end{equation}
where
\begin{equation}
U_{H_f}\equiv\left(\frac{1+\epsilon H_f}{1-\epsilon^2 H_f}\right)\cos\theta_f'\cos\gamma,
\end{equation}
and
\begin{equation}
V_{H_f}\equiv\frac{1-H_f}{1-\epsilon^2 H_f}.
\end{equation}
Here the main uncertainty is associated with the determination of $H_f$,
which relies on (\ref{SU3-1}). In Subsection~\ref{ssec:SU3-H}, we have 
a closer look at the corresponding $SU(3)$-breaking corrections, and give
numerical results for the extraction of the $H_f$ from the untagged 
observables. On the other hand, thanks to the $\epsilon$ terms in 
(\ref{H-def}), the impact of corrections to (\ref{SU3-2}) is tiny.

Another useful quantity is offered by the direct CP asymmetry
\begin{equation}\label{af-ADdet}
\hat A_{\rm D}^{f'}=\frac{2 a_f'\sin\theta_f'\sin\gamma}{1-2a_f'\cos\theta_f'\cos\gamma
+a_f'^2},
\end{equation}
which can be extracted from a rate difference; it takes the same form as
(\ref{t-dep-asym}) for $t=0$. In analogy to $H_f$, also $\hat A_{\rm D}^{f'}$ 
allows us to determine $a_f'$ as a function of $\theta_f'$. To this end, we may 
again use (\ref{af-Hdet}), with the following replacements:
\begin{equation}\label{UVD}
U_{H_f} \,\to\, U_{\hat A_{\rm D}^{f'}}\equiv \cos\theta_f'\cos\gamma+
\frac{\sin\theta_f'\sin\gamma}{\hat A_{\rm D}^{f'}}, \quad
V_{H_f} \,\to\, V_{\hat A_{\rm D}^{f'}}\equiv 1.
\end{equation}
It should be emphasized that the corresponding curve in the $\theta_f'$--$a_f'$ 
plane is theoretically clean, whereas that described by (\ref{af-Hdet}) is
affected in particular by the $SU(3)$-breaking effects entering the determination 
of $H_f$.

The intersection of the $H_f$ and $\hat A_{\rm D}^{f'}$ contours allows us then
to extract $a_f'$ and $\theta_f'$ from the data. Finally, applying (\ref{SU3-2})
and the results derived in Section~\ref{ssec:phis-impact}, we can include the 
penguin effects in the determination of the $B^0_s$--$\bar B^0_s$ mixing phase. 
Let us first illustrate this method in the next section by discussing a numerical 
example before giving a detailed discussion of the relevant $SU(3)$-breaking 
effects in Section~\ref{sec:uncert}.

\section{A Numerical Example}\label{sec:illu}
For the illustration of the strategy discussed above, we assume 
$\gamma=65^\circ$, and hadronic parameters given by $a_f'=0.4$ and 
$\theta_f'=220^\circ$, yielding the observables $H_f=1.44$ and 
$\hat A_{\rm D}^{f'}=-0.33$. These input values are consistent with the 
ranges of the $B^0_d\to J/\psi \pi^0$ parameters $a'\in[0.15,0.67]$ and 
$\theta'\in[174^\circ,213^\circ]$ found in Ref.~\cite{FFJM}; we expect
a picture for $a_f'$ and $\theta_f'$ that is similar to the one for their 
$B^0_d\to J/\psi \pi^0$ counterparts.

\begin{figure}
   \centering
   \includegraphics[width=8.0truecm]{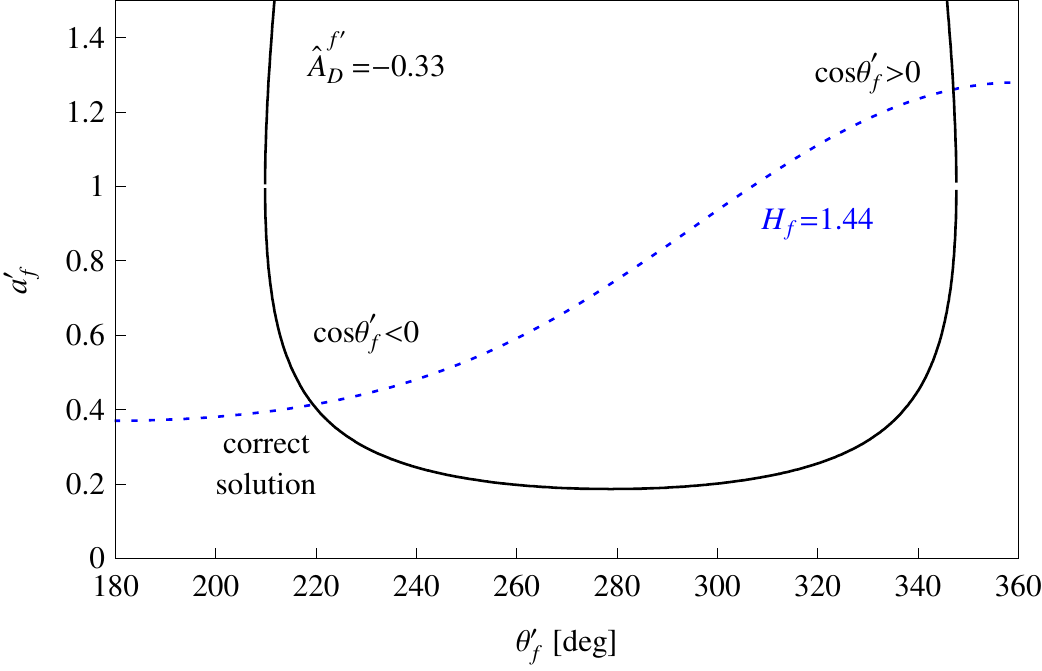} 
   \caption{Illustration of the contours in the $\theta_f'$--$a_f'$ plane for 
   an example, as discussed in the text.}\label{fig:6}
\end{figure}

In Fig.~\ref{fig:6}, we show the contours in the  $\theta_f'$--$a_f'$ plane
that arise in this example. We observe that a twofold solution emerges
for $(\theta_f',a_f')$, which can be resolved through the sign of
$\cos\theta_f'$. Theoretically, we expect a negative value of this quantity, 
which is also supported by the $B^0_d\to J/\psi \pi^0$ data. In order to 
resolve this ambiguity experimentally, we need an additional observable,
which would be provided by mixing-induced CP violation. Since the
$B^0_s\to J/\psi [\to\ell^+\ell^-]\bar K^{*0}[\to\pi^+K^-]$ processes have
flavour-specific final states, they do not show this phenomenon. On the
other hand, mixing-induced CP violation would arise in 
$B^0_s\to J/\psi [\to\ell^+\ell^-]\bar K^{*0}[\to\pi^0K_{\rm S,L}]$ modes,  
in analogy to $B^0_d\to J/\psi [\to\ell^+\ell^-]\bar K^{*0}[\to\pi^0K_{\rm S,L}]$ 
processes \cite{DDF}. Unfortunately,  it is essentially impossible to study the 
corresponding experimental signatures in a hadronic environment, i.e.\ at the 
Tevatron or LHC.

\begin{figure}
   \centering
   \includegraphics[width=8.0truecm]{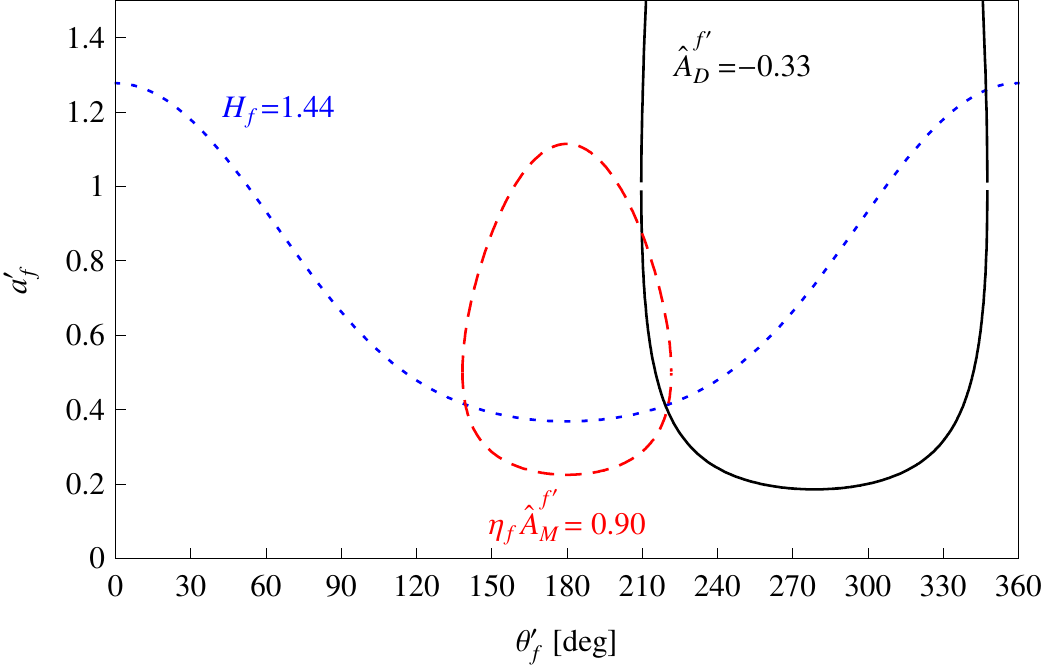} 
   \caption{Illustration of the resolution of the twofold ambiguity in 
   Fig.~\ref{fig:6} through the mixing-induced CP violation in 
   $B^0_d\to J/\psi\rho^0$.}\label{fig:7}
\end{figure}

However, we may alternatively use the $B^0_d\to J/\psi \rho^0$ 
channel \cite{RF-ang}, which can be obtained
from $B^0_s\to J/\psi \bar K^{*0}$ by replacing the strange spectator quark
through a down quark, as can be seen in Fig.~\ref{fig:5}. In this case, 
the final state is an admixture of different CP eigenstates, in analogy to
$B^0_s\to J/\psi \phi$, and we can extract the following mixing-induced
CP asymmetry from the time-dependent angular distribution:
\begin{equation}\label{AM-psirho0}
 \hat A_{\rm M}^{f '}=+\eta_f\left[\frac{\sin\phi_d-2a_f'\cos\theta_f'
 \sin(\phi_d+\gamma)+a_f'^2\sin(\phi_d+2\gamma)}{1- 2a_f'\cos\theta_f'
 \cos\gamma+a_f'^2}\right],
\end{equation}
where $\eta_f$ is the CP eigenvalue of the final-state configuration
$f$, i.e., $\eta_0$, $\eta_\parallel=+1$ and $\eta_\perp=-1$, whereas
$\phi_d=(42.4^{+3.4}_{-1.7})^\circ$  denotes the $B^0_d$--$\bar B^0_d$ 
mixing phase \cite{FFJM}; for simplicity, we have
also denoted the  $B^0_d\to J/\psi \rho^0$ hadronic parameters by
$a_f'$ and $\theta_f'$, as we expect them to be approximately equal to those
of $B^0_s\to J/\psi \bar K^{*0}$ thanks to the $SU(3)$ flavour symmetry. 
Using (\ref{af-Hdet}) with the replacements
\begin{equation}
U_{H_f} \,\to\, U_{\hat A_{\rm M}^{f'}} \equiv \left[\frac{\sin(\phi_d+\gamma)-
 \hat A_{\rm M}^{f '}\cos\gamma}{\sin(\phi_d+2\gamma)- \hat A_{\rm M}^{f '}}
 \right]\cos\theta',
\end{equation}
\begin{equation}
 V_{H_f} \,\to\, V_{\hat A_{\rm M}^{f'}}\equiv 
 \frac{\sin\phi_d-\hat A_{\rm M}^{f'}}{\sin(\phi_d+2\gamma)- \hat A_{\rm M}^{f '}},
\end{equation}
the measurement of the mixing-induced CP asymmetry $ \hat A_{\rm M}^{f '}$
allows us to fix another contour in the $\theta_f'$--$a_f'$ plane. If we consider
the example given above with $\phi_d=42.4^\circ$, we obtain 
$ \eta_f\hat A_{\rm M}^{f '}=0.90$, which
results in the contours shown in Fig.~\ref{fig:7}. We see that the twofold ambiguity
in the determination of the hadronic parameters can now be resolved, 
thereby leaving us with our input values.

Since the width of the $\rho^0$ is three-times larger than that of the $\bar K^{*0}$, 
the  $B^0_s\to J/\psi \bar K^{*0}$ control channel 
should be experimentally better accessible than $B^0_d\to J/\psi \rho^0$.
Moreover, if we neglect $SU(3)$-breaking effects due to the different spectator
quarks, we expect the simple relation 
$\mbox{BR}(B^0_s\to J/\psi \bar K^{*0})\sim 2\times 
\mbox{BR}(B^0_d\to J/\psi \rho^0)=(4.6\pm0.4)\times10^{-5}$ \cite{HFAG}.
However, already a rather crude measurement of the mixing-induced 
CP-violating observables of  $B^0_d\to J/\psi \rho^0$
would be sufficient to resolve the ambiguity in the extraction of $a_f'$ and $\theta_f'$.
In particular, the expected negative value of $\cos\theta_f'$ would be indicated
by values of $\eta_f\hat A_{\rm M}^{f '}$ that are larger than $\sin\phi_d=0.67$.
Such a pattern emerges actually in the measurement of the mixing-induced
CP violation of $B^0_d \to J/\psi \pi^0$.

\begin{figure}
   \centering
 \includegraphics[width=8.0truecm]{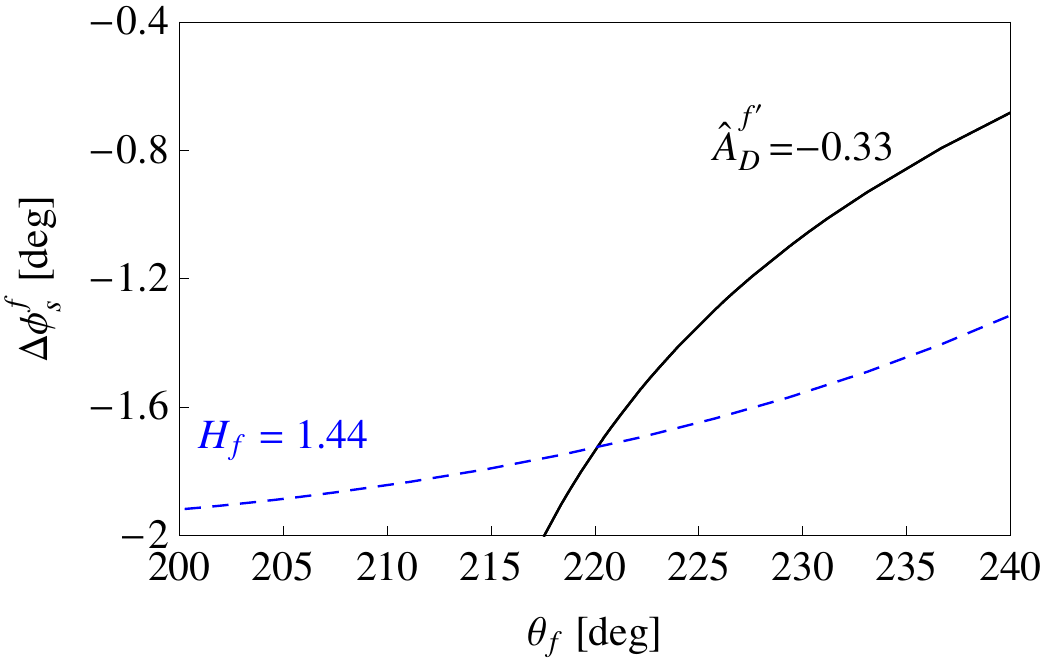} 
   \caption{Situation in the $\theta_f$--$\Delta\phi^f_s$ plane for the example
   in Figs.\ \ref{fig:6} and \ref{fig:7}.}\label{fig:8}
\end{figure}

In Fig.~\ref{fig:8}, we convert the contours in the $\theta_f'$--$a_f'$ plane
into the $\theta_f$--$\Delta\phi^f_s$ space by means of (\ref{sDelphis})--(\ref{tDelphis})
and (\ref{SU3-2}). We observe that, in this specific example, the shift of the 
$B^0_s$ --$\bar B^0_s$ mixing phase through the penguin effects
is given by $\Delta\phi^f_s=-1.7^\circ$. If we assume the SM, the 
mixing-induced CP asymmetries of $B^0_s\to J/\psi \phi$ 
represented by (\ref{AM-pen-expr}) would be given by
$\eta_f\hat A_{\rm M}^f=-6.7\%$, which is about twice as large as the
SM value. At LHCb, such CP asymmetries could be detected with about 
$4\,\sigma$ significance after collecting $2\,\mbox{fb}^{-1}$ of data, 
corresponding to one nominal year of operation, and with about $20\,\sigma$
at an upgrade of this experiment with $100\,\mbox{fb}^{-1}$ integrated
luminosity. However, without the control of the hadronic penguin effects through 
a simultaneous analysis of the $B^0_s\to J/\psi \bar K^{*0}$ channel as 
proposed above, this result would be misinterpreted as a signal of physics 
beyond the SM. In this context it is important to emphasize that we expect 
$\phi_s^{\rm SM}$ and $\Delta\phi^f_s$ to have the same negative sign, thereby 
leading to constructive interference. In the opposite case, i.e.\ with a positive 
value of $\Delta\phi^f_s$, the SM picture of expecting vanishingly small 
CP violation in $B^0_s\to J/\psi \phi$ would be much more robust with 
respect to the hadronic penguin uncertainties. It cannot be excluded that
the hadronic penguin effects are actually more significant than in our example, and
could lead to $\eta_f\hat A_{\rm M}^f\sim-10\%$. This feature is fully
supported by the picture emerging from the current $B_d^0\to J/\psi \pi^0$ data
\cite{FFJM}.

\begin{figure}
   \centering
   \includegraphics[width=8.0truecm]{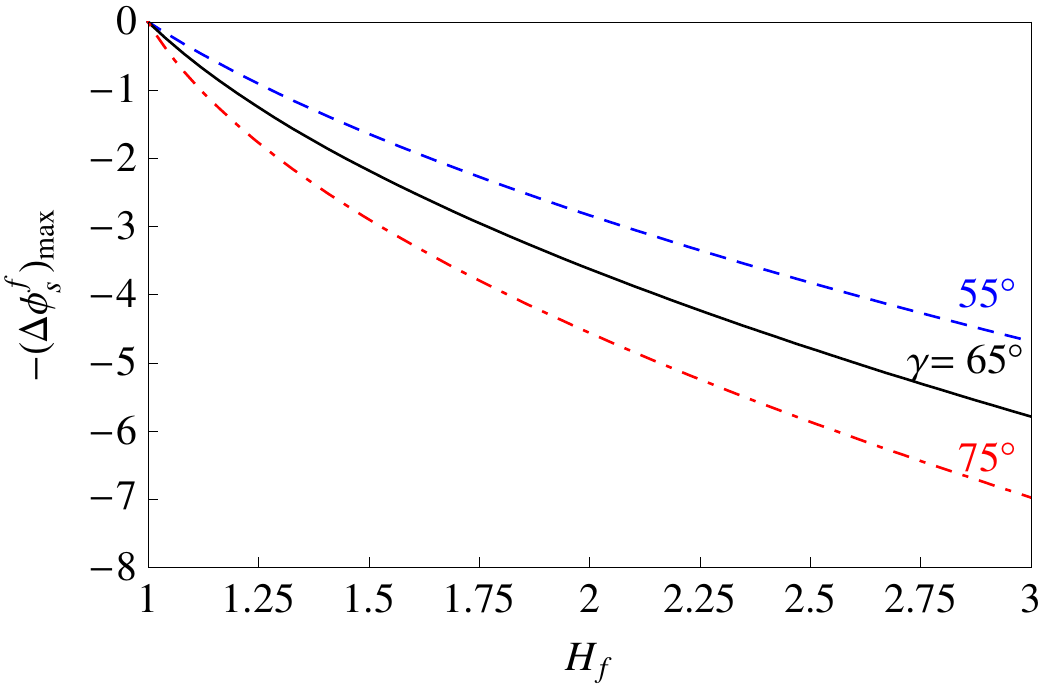} 
   \caption{The maximal shift of $\phi_s$ as function of $H_f$ for various
   values of $\gamma$.}\label{fig:9}
\end{figure}

In view of these findings, it would be very desirable to search for the
$B^0_s\to J/\psi \bar K^{*0}$ decay at the Tevatron. Already information
on $H_f$ would allow us to put first valuable constraints on the shift
$\Delta\phi_s^f$. As we have shown in Fig.~\ref{fig:9}, these observables
will put a first upper bound on $\Delta\phi_s^f$. Once direct
CP violation in the $B^0_s\to J/\psi \bar K^{*0}$ angular distribution is 
measured, $\Delta\phi_s^f$ can be fully pinned down, as we have shown 
above.

\boldmath
\section{$SU(3)$-Breaking Effects} \label{sec:uncert}
\unboldmath
\boldmath
\subsection{General Remarks and $\omega$--$\phi$ Mixing}
\unboldmath
The main theoretical uncertainty of the strategy proposed in the present
paper is related to $SU(3)$-breaking effects which affect the relations 
in (\ref{SU3-1}) and (\ref{SU3-2}). The following discussion of $SU(3)$ breaking 
is parallel to our previous investigation \cite{FFJM}; however,  here we deal with 
vector meson final states. On the one hand this simplifies the discussion
since $SU(3)$-breaking effects in the vector meson octett seem to be smaller;  
on the other hand, we have a $\phi$ state which is believed to be to a good 
approxmation an $\bar{s}  s$ state and hence a superposition of $SU(3)$ 
eigenstates.

The $SU(3)$ nonet contains three neutral, non-strange states. Assuming isospin 
to be a good symmetry, one of these states is the neutral $I=1$  $\rho$ meson with the 
quark decomposition $\rho_0 = (u\bar{u} - d \bar{d}) / \sqrt{2}$. The other two states are
isosinglets and are given by $\phi_0 = (u\bar{u} + d \bar{d} + s\bar{s} ) / \sqrt{3}$, which is an 
$SU(3)$ singlet, and  $\phi_8 = (u\bar{u} + d \bar{d} - 2 s\bar{s} ) / 4$ belonging to the 
$SU(3)$ octet. In case of unbroken $SU(3)$ symmetry, the amplitudes  for 
processes involving the members of the octet are related to one another, 
while the singlet remains separate. 

However, there is good indiction that the physical $\phi$ state is to a good approximation 
a pure $s \bar{s}$ state and hence a superposition of the singlet $\phi_0$ and the octet 
$\phi_8$. The orthogonal state $\omega =    (u\bar{u} + d \bar{d}) / \sqrt{2}$ is the isoscalar 
$\omega$ meson. Up to small mixing between $\omega$ and $\phi$, which has been 
discussed recently in the context of $B$ decays in Ref.~\cite{Gronau:2008kk}, and 
which is too small to be relevant here,  these are the (strong) mass eigenstates. 

The way we apply the $SU(3)$ symmetry is to assume that the matrix 
elements of $\phi =  s\bar{s} $ are related to the corresponding matrix elements 
of the members of the octet. In other words, we shall assume that the form 
factors of the $B^0_s \to \phi$ transition are the same as the ones for the 
$B^0_s \to \bar K^{*0}$ decay. Strictly speaking, this goes beyond  the $SU(3)$ 
symmetry assumption  since we relate octet and singlet components. 
Lacking any detailed information on the quality of such an asumption we 
have to rely, e.g., on QCD sum rule estimates which indicate that the 
strong dynamics in the $\phi =  (s\bar{s})_{S=1} $ state are very similar 
to $K^{*0} =  (d\bar{s})_{S=1} $; in fact, we shall rely on QCD sum rules 
in Section~\ref{ssec:SU3-H} to discuss the deviations  from our assumption. 
In this context, it should be emphasized again that we have also to neglect 
penguin annihilation and exchange topologies, which can be probed through 
the $B^0_d\to J/\psi \phi$ decay. 

In a recent paper \cite{GR-psi-phi}, it is argued in detail that the relations
between $B^0_s\to J/\psi \phi$ and $B^0_d\to J/\psi K^{*0}$ following from
flavour symmetry \cite{DDF-2} are likely to be quite reliable, so that using strong
phase information from the $B^0_d\to J/\psi K^{*0}$ channel in the analysis 
of $B^0_s\to J/\psi \phi$ is justified. Also here we have to assume that the 
matrix elements of $\phi =  s\bar{s} $ are related to the corresponding matrix 
elements of the members of the octet. 

It is very difficult to get a reliable  estimate of the $SU(3)$ breaking for the non-leptonic decays at hand. It is known from the corresponding processes with pseudoscalar final states that the decays with $J/\psi$ in the final state are dominated by non-factorizable contributions; it is not even clear how to factorize the penguin 
contributions in the decays we are considering. However, in the case 
of $B_{(s)}\to \pi\pi,\pi K,KK$ decays, we encounter sizeable non-factorizable effects,
whereas the data do not indicate large $SU(3)$-breaking effects of this kind 
\cite{RF-BpiKKK}. In particular, considering the counterpart of the $H_f$ quantities
introduced in the present paper for the $B^0_s\to K^+K^-$, $B^0_d\to\pi^+\pi^-$
system, a calculation of the relevant form-factor ratio by means of QCD sum rule 
techniques \cite{KMM} yields good agreement with the current data that would
be spoiled by large non-factorizable, $SU(3)$-breaking effects. 

This empirical behaviour gives us conficence that our estimate of the 
$SU(3)$-breaking effects for the extraction of the $H_f$ from the data 
given in the next subsection, which relies on a QCD sum rule analysis 
of the relevant form factors as well, describes the leading corrections.

\boldmath
\subsection{$SU(3)$ Breaking in the Extraction of $H_f$}\label{ssec:SU3-H}
\unboldmath
In order to calculate the $SU(3)$-breaking corrections to the amplitude ratios
$|{\cal A}_f/{\cal A}_f'|$ that are required for the extraction of the $H_f$ from
the data (see (\ref{H-def})), we apply the formulae given in Ref.~\cite{DDF}. The linear
polarization amplitudes of the $B^0_s\to J/\psi \phi$ channel at time $t=0$ can be 
written as
\begin{equation}\label{obs-def}
\begin{split}
A_0(0) &= -xa - (x^2- 1)b\ \\
A_\parallel (0) &= \sqrt 2 a  \\
A_\perp (0) &= \sqrt{2(x^2 - 1)} \ c
\end{split}
\end{equation}
with
\begin{equation}\label{x-def}
x\equiv\frac{p_{J/\psi}\cdot p_{\phi}}{m_{J/\psi}m_{\phi}}=
\frac{m_{B_s}^2-m_{J/\psi}^2-m_{\phi}^2}{2m_{J/\psi}m_{\phi}},
\end{equation}
where the ``factorized" contributions are given by 
\begin{equation}
 \begin{split}
a_{\rm fact} &= \frac{G_{\rm F}}{\sqrt 2}
\lambda_c^{(s)}
\bigl( \mathcal C_1^\text{eff}(\mu) + \mathcal C_5^\text{eff} (\mu) \bigr)  A_1^{\rm fact}\ , \\
b_{\rm fact} &= \frac{G_{\rm F}}{\sqrt 2}\lambda_c^{(s)}
\bigl( \mathcal C_1^\text{eff}(\mu) + \mathcal C_5^\text{eff} (\mu) \bigr)
B_1^{\rm fact},\\
c_{\rm fact} &= \frac{G_{\rm F}}{\sqrt 2}\lambda_c^{(s)}
\bigl( \mathcal C_1^\text{eff}(\mu) + \mathcal C_5^\text{eff} (\mu) \bigr)
 C_1^{\rm fact}.
\end{split}
\end{equation}
Here we have neglected the doubly Cabibbo-suppressed penguin corrections, 
as our target are the overall amplitudes ${\cal A}_f$; $G_{\rm F}$ is Fermi's
constant, $\lambda_c^{(s)}$ the CKM factor introduced after (\ref{Bspsiphi-ampl}), 
and the $\mathcal C_i^\text{eff}(\mu)$ are the ``effective" Wilson coefficient 
functions introduced in Ref.~\cite{DDF}. Moreover, we have 
\begin{align}\label{FF-1}
  A_1^{\rm fact} &= - f_{J/\psi} m_{J/\psi} (m_{B_s} + m_\phi ) A_1^{B_s \phi} (m_{J/\psi}^2) \ ,\notag \\
  B_1^{\rm fact} &= 2 \ \frac{f_{J/\psi} m_{J/\psi}^2 m_\phi}{m_{B_s} + m_\phi}\ A_2^{B_s \phi}(m_{J/\psi}^2)\ ,\\
  C_1^{\rm fact} &= 2 \ \frac{f_{J/\psi} m_{J/\psi}^2 m_\phi}{m_{B_s}+m_\phi} \ V^{B_s\phi}(m_{J/\psi}^2), \notag
\end{align}
where  $A_{1,2}^{B_s \phi}(q^2)$ and $V^{B_s \phi}(q^2)$ are the form
factors of the quark-current matrix elements of the $B_s \to \phi$ transition,
with $q$ denoting the momentum transferred by the quark current. 
In the case of the $B^0_s\to J/\psi \bar K^{*0}$ channel, we need correspondingly
the $B_s\to \bar K^{*0}$ transition form factors, and have to replace 
$\phi \to \bar K^{*0}$ in (\ref{x-def}) and (\ref{FF-1}).

\begin{table}
\begin{center}
\begin{tabular}{|c|c|c|}
\hline 
& $V=\phi$ & $V= \bar K^{*}$ \\
\hline
$A_1^{B_s\to V}(m_{J/\psi}^2) $ &   $ 0.42 \pm 0.06$ &   $0.33 \pm 0.05 $       
\\
$A_2^{B_s\to V}(m_{J/\psi}^2)$ &   $0.38 \pm 0.06 $ &    $0.32 \pm 0.05 $      
\\
$V^{B_s\to V}(m_{J/\psi}^2)$ & 0.82 $ \pm 0.12 $  &    $0.62 \pm 0.09 $    \\
\hline
\end{tabular}
\end{center}
\caption{Collection of the relevant $B_s\to V$ form factors at 
$q^2=m_{J/\psi}^2$, using
the results of Ref.~\cite{FF-BZ} and assuming an uncertainty of 
$15\%$.}\label{tab:1}
\end{table}

An analysis of these form factors was performed in Ref.~\cite{FF-BZ}. Light 
cone QCD sum rules allow an estimate of the values of the form factors at 
$q^2 = 0$. In order to obtain the value of the form factor at a different 
$q^2$, such as $q^2=m_{J/\psi}^2$ as in (\ref{FF-1}), we have to 
make an extrapolation using some parametrization of the form factor.
If we use the functional forms suggested in Ref.~\cite{FF-BZ} and assume
an uncertainty of $15\%$, we obtain the form factors at $q^2= m_{J/\psi}^2$
collected in Table~\ref{tab:1}, and the following $SU(3)$-breaking ratios:
\begin{equation}\label{FFBZ}
\begin{split}
 \frac{A_1^{\BsKdec}(\MJpsi) }{A_1^{\Bphidec}(\MJpsi)} &= 0.78\pm 0.08 \ , \\
\frac{A_2^{\BsKdec}(\MJpsi) }{A_2^{\Bphidec}(\MJpsi)} &= 0.84\pm 0.07 \ , \\
\frac{V^{\BsKdec}(\MJpsi) }{V^{\Bphidec}(\MJpsi)} &= 0.76\pm 0.15 \ . \\
\end{split}
\end{equation} 
Using then \eqref{obs-def}, we obtain the following numerical results, which 
allow the extraction of the $H_f$ from the untagged rates with the help
of (\ref{H-def}):
\begin{equation}
 \begin{split}
 \left|\frac{{\cal A}'_0}{{\cal A}_0}\right|^2
&= 0.42 \pm  0.27\ , \\
 \left|\frac{{\cal A}'_\parallel}{{\cal A}_\parallel}\right|^2
&= 0.70 \pm 0.29\ , \\
\left|\frac{{\cal A}'_\perp}{{\cal A}_\perp}\right|^2
&= 0.38 \pm 0.16 \ .
\end{split}
\end{equation}
Note that in order to calculate $|{\cal A}'_0/{\cal A}_0|^2$, we need the
$A_{1,2}^{B_s\to V}(m_{J/\psi}^2)$ form factors given in Table~\ref{tab:1}.

\boldmath
\subsection{$SU(3)$ Breaking in $a_f'=a_f$ and $\theta_f'=\theta_f$}\label{ssec:a-theta}
\unboldmath
If we use the $B^0_s\to J/\psi \bar K^{*0}$ observables as discussed in 
Section~\ref{sec:contr}, we can extract $a_f'$ and 
$\theta_f'$ from the data. Since their $B^0_s\to J/\psi \phi$ counterparts $a_f$ 
and $\theta_f$ enter in $H_f$ in combination with the tiny parameter 
$\epsilon$, this determination is essentially unaffected by corrections to 
(\ref{SU3-2}); the main corrections enter through the value of $H_f$, which 
requires the amplitude ratios $|{\cal A}_f/{\cal A}_f'|$, with the $SU(3)$-breaking
corrections estimated in the previous subsection. 

When calculating the shifts $\Delta\phi_s^f$, we have to use 
the relations in (\ref{SU3-2}). However, one has to keep in mind that 
sizable non-factorizable effects could induce $SU(3)$-breaking corrections. 
Their impact on the determination of $\Delta\phi_s^f$ can be easily inferred from 
(\ref{tDelphis}). Neglecting terms of order $\epsilon^2$, we have a linear 
dependence on $a_f\cos\theta_f$. Consequently, corrections to the left-hand 
side of (\ref{SU3-2}) propagate linearly, while $SU(3)$-breaking effects 
in the strong phases will generally lead to an asymmetric uncertainty for 
$\Delta\phi_s^f$.

In the analysis of the $B^0_d\to J/\psi \pi^0$ data in Ref.~\cite{FFJM}, the
impact of $SU(3)$-breaking corrections was explored by setting
$a=\xi a'$ and uncorrelating the strong phases $\theta$ and $\theta'$
of the $B^0_d\to J/\psi K^0$ and $B^0_d\to J/\psi\pi^0$ decays, respectively.  
Even when allowing for $\xi\in[0.5,1.5]$ and $\theta,\theta'\in[90,270]^\circ$ in 
the corresponding fit, and using a $50\%$ increased error for the relevant
form-factor ratio to explore the impact of dramatic non-factorizable, $SU(3)$-breaking
contributions to $|{\cal A}/{\cal A}'|$, the picture emerging from the global fit is not 
significantly changed. To be specific, $\Delta\phi_d\in[-6.7,0.0]^\circ$ arises 
when allowing for such large $SU(3)$-breaking corrections, whereas 
$\Delta\phi_d\in[-3.9,-0.8]^\circ$ in the case with $\xi=1$ and $\theta=\theta'$. 
We expect a similar situation for $\Delta \phi_s^f$.

\boldmath
\subsection{Internal Consistency Checks of $SU(3)$}
\unboldmath
The advantage of $B$ decays into two vector mesons is that many more
observables are offered by the angular distribution of their decay products
than in the case of $B\to PP$ or $B\to PV$  decays ($P$ and $V$ denote
generically pseudoscalar and vector mesons, respectively). This comment applies
also to the decays considered in the present paper, and allows us to
perform internal consistency checks of the $SU(3)$ flavour symmetry.

\begin{figure}
   \centering
   \includegraphics[width=8.0truecm]{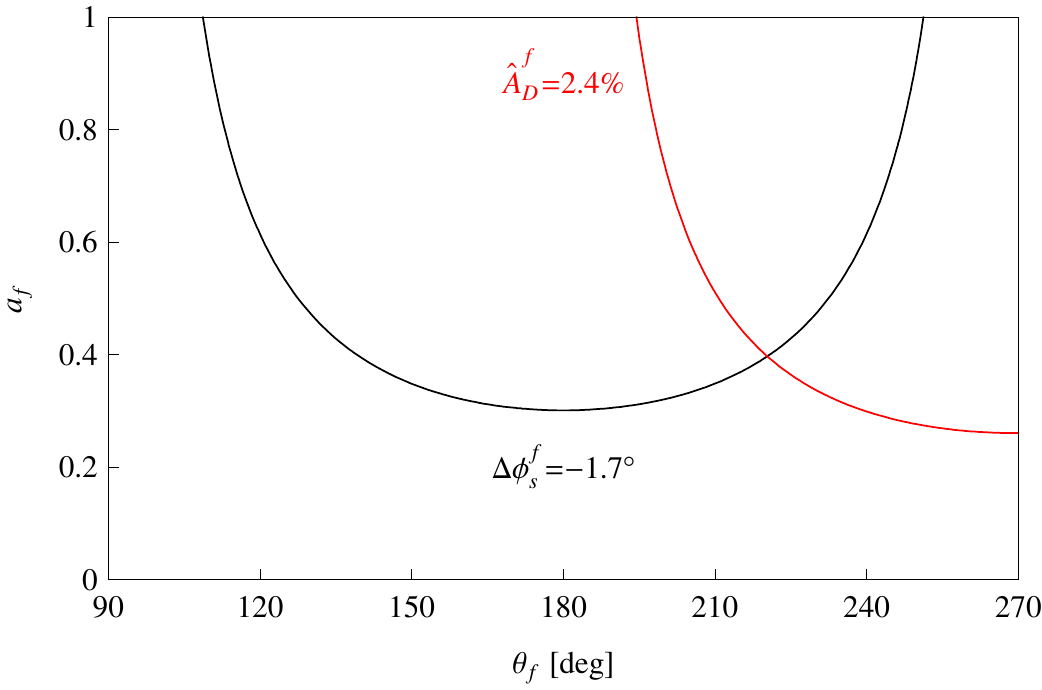} 
   \caption{The extraction of $a_f$ and $\theta_f$ from $\Delta\phi_s^f$ and
   the direct CP asymmetry $\hat A_{\rm D}^f$ for internal consistency
   checks of the $SU(3)$ flavour symmetry, as described in the text. 
   The example corresponds to $a_f=0.4$ and $\theta_f=220^\circ$ with
   $\gamma=65^\circ$, as in the previous numerical illustrations.}\label{fig:10}
\end{figure}

A very first internal test follows from a comparison of the 
different values of the $B^0_s$--$\bar B^0_s$ mixing phase $\phi_s$ 
following from the three polarization states $f\in\{0,\parallel,\perp\}$.
Obviously, these values should agree with one another. 
In fact, even more quantitative tests of $SU(3)$ breaking can be performed. 
The point is that we may choose one of the three linear polarization states
to extract $\phi_s$ from (\ref{AM-pen-expr}), taking the shift $\Delta\phi_s^f$ 
through the penguin effects into account. Using then the 
$B^0_s\to J/\psi \phi$ observables $\hat A_{\rm M}^f$ and $\hat A_{\rm M}^f$ 
of the remaining two polarization states, the knowledge of $\phi_s$ allows us 
to extract the corresponding shifts $\Delta\phi_s^f$ from  (\ref{AM-pen-expr}).
With the help of (\ref{tDelphis}), we can then convert the values of the 
$\Delta\phi_s^f$ into contours in the $\theta_f$--$a_f$ plane. To this
end, we have simply to make the following replacements in 
(\ref{af-Hdet}):
\begin{equation}
U_{H_f}\to U_{\Delta\phi_s^f}\equiv 
\left(\frac{\sin\gamma-\cos\gamma\tan\Delta\phi_s^f}{\cos2\gamma\tan\Delta\phi_s^f-
\sin2\gamma}\right)\cos\theta_f,
\end{equation}
\begin{equation}
V_{H_f}\to V_{\Delta\phi_s^f}\equiv \frac{\tan\Delta\phi_s^f}{\cos2\gamma
\tan\Delta\phi_s^f-\sin2\gamma},
\end{equation}
as well as $a_f'\to \epsilon a_f$. Moreover, if we replace, in addition to the
latter substitution,  $\theta_f'\to 180^\circ+\theta_f$ and $\hat A_{\rm D}^{f'}
\to \hat A_{\rm D}^f$ in  (\ref{UVD}), the direct CP asymmetry in 
$B^0_s\to J/\psi \phi$ can be converted into a contour in the 
$\theta_f$--$a_f$ plane as well. It should be stressed that these constructions 
are valid exactly. In Fig.~\ref{fig:10}, we illustrate 
how this works by considering again the numerical example specified in 
Section~\ref{sec:illu}.

The values of the hadronic $B^0_s\to J/\psi\phi$ parameters $a_f$ and $\theta_f$ 
allow us then to perform an internal consistency check of the $SU(3)$ flavour 
symmetry by comparing with the values of $a_f'$ and $\theta_f'$ following 
from the $B^0_s\to J/\psi \bar K^{*0}$ strategy proposed in Section~\ref{sec:contr}. 
Another test is offered by the following relations:
\begin{equation}
 \hat A_{\rm D}^f=-\epsilon H_f  \hat A_{\rm D}^{f'},
\end{equation}
which rely on (\ref{SU3-2}). Needless to note, the practical usefulness of these consistency checks depends on the values of the observables that will eventually 
be measured by LHCb. We strongly encourage detailed feasibility studies and 
look forward to confronting these considerations with real data soon.

\section{Conclusions}\label{sec:concl}
Studies of CP-violating effects in the time-dependent angular distribution of
$B^0_s\to J/\psi[\to\ell^+\ell^-] \phi[\to K^+K^-]$ processes 
have recently received considerable attention
in view of first tagged measurements at the Tevatron, and are a central target
of the LHCb experiment which will soon start taking data. We have pointed
out that hadronic effects, which are due to doubly Cabibbo-suppressed
penguin contributions that are usually neglected, could induce 
mixing-induced CP-violating effects as large as ${\cal O}(-10\%)$. Without the control
of these penguin contributions, which cannot be calcuated reliably from
QCD, such CP-violating effects, which can be detected with excellent significance by 
LHCb, would be misinterpreted as CP-violating NP contributions to 
$B^0_s$--$\bar B^0_s$ mixing. 

In the present paper, we have proposed a strategy to include these contributions 
with the help of a measurement of the angular distribution of the 
$B^0_s\to J/\psi[\to \ell^+\ell^-] \bar K^{\ast 0}[\to\pi^+K^-]$ decay products,
and have illustrated this by means of a numerical example. We strongly 
suggest a search for this control channel at the Tevatron in order to obtain 
first constraints on the penguin effects in the $B^0_s\to J/\psi \phi$ analysis. 
The tremendous accuracy that can be achieved at LHCb and a possible future 
upgrade of this experiment makes it mandatory to include these penguin 
contributions.

\vspace*{0.5truecm}

\paragraph{Acknowledgements:} \noindent
S.F. and T.M. acknowledge the support by the German Ministry of 
Research (BMBF, Contract No.~05HT6PSA), and R.F. would like to
thank Guido Martinelli for the hospitality at the Universit\`a di Roma ``La Sapienza".

\newpage

\section*{Appendix}
\appendix
\boldmath
\section{Time-dependent Angular Distributions of 
$B^0_s\to J/\psi \bar K^{\ast0}$ and CP Conjugates}\label{app-a}
\unboldmath
Following Ref.~\cite{DDF}, we introduce the following set of trigonometric functions:
\begin{equation}
\begin{array}{rcl}
f_1 & = & 2\, \cos^2\psi\, (1 - \sin^2\theta \,\cos^2\varphi) 
 \\
f_2 & = & \sin^2\psi\, (1 - \sin^2\theta \,\sin^2\varphi)
 \\
f_3 & =  & \sin^2\psi \,\sin^2\theta
 \\
f_4 & = & \sin^2\psi\,\sin2\theta\, \sin\varphi
 \\
f_5 & = & (1/\sqrt{2}) \,\sin2\psi\,\sin^2\theta \sin2 \varphi
 \\
f_6 & = & (1/\sqrt{2}) \,\sin2\psi\, \sin2 \theta\, \cos\varphi.
\end{array}
\end{equation}
If we use the notation $A_f\equiv A(B_s^0\to (J/\psi \bar K^{*0})_f)$ for the 
unevolved amplitude in (\ref{ampl-d}) and $\bar A_f$ for its CP conjugate, 
we obtain
\begin{eqnarray}
\lefteqn{\frac{d^3 \Gamma [B^0_s(t) \to J/\psi (\to \ell^+ \ell^-) \bar K^{\ast0} 
(\to \pi^+ K^-)]}
{d \cos \theta~d \varphi~d \cos \psi}
=\frac{9}{64 \pi}\left[\cosh(\Delta\Gamma_st/2)+\cos(\Delta M_st)\right]
e^{-\Gamma_s t}}\nonumber\\
&&\times \left\{ f_1 |A_0|^2   + f_2 |A_{\|}|^2 + f_3 |A_{\perp}|^2 
 - f_4 {\mbox  Im}\,(A_{\|}^{\ast} A_{\perp})  
 + f_5  {\mbox Re}\,(A_0^{\ast} A_{\|})
 + f_6  {\mbox Im}\,(A_0^{\ast} A_{\perp}) \right\}\nonumber\\
\end{eqnarray}
\begin{eqnarray}
\lefteqn{\frac{d^3 \Gamma [\bar B^0_s(t) \to J/\psi (\to \ell^+ \ell^-) 
K^{\ast0} (\to \pi^-K^+)]}
{d \cos \theta~d \varphi~d \cos \psi}=
\frac{9}{64 \pi}\left[\cosh(\Delta\Gamma_st/2)+\cos(\Delta M_st)\right]
e^{-\Gamma_s t}}\nonumber\\
&&\times\left\{ f_1 |\bar A_0|^2   + f_2 |\bar A_{\|}|^2 + f_3 |\bar A_{\perp}|^2 
 + f_4\, {\mbox  Im}\,(\bar A_{\|}^{\ast} \bar A_{\perp})  
 + f_5\, {\mbox Re}\,(\bar A_0^{\ast} \bar A_{\|})
 -  f_6\, {\mbox Im}\,(\bar A_0^{\ast}\bar A_{\perp}) \right\}\nonumber\\
\end{eqnarray}
\begin{eqnarray}
\lefteqn{\frac{d^3 \Gamma [B^0_s(t) \to J/\psi (\to \ell^+ \ell^-) K^{\ast0} 
(\to \pi^- K^+)]}{d \cos \theta~d \varphi~d \cos \psi}=
\frac{9}{64 \pi}\left[\cosh(\Delta\Gamma_st/2)-\cos(\Delta M_st)\right]
e^{-\Gamma_s t}}\nonumber\\
&&\times \left\{ f_1 |\bar A_0|^2  + f_2 |\bar A_{\|}|^2 + f_3 |\bar A_{\perp}|^2 
 + f_4\, {\mbox  Im}\,(\bar A_{\|}^{\ast} \bar A_{\perp})  
 + f_5\, {\mbox Re}\,(\bar A_0^{\ast} \bar A_{\|})
 -  f_6\, {\mbox Im}\,(\bar A_0^{\ast} \bar A_{\perp}) \right\}\nonumber\\
\end{eqnarray}
\begin{eqnarray}
\lefteqn{\frac{d^3 \Gamma [\bar B_s(t) \to J/\psi (\to \ell^+ \ell^-) \bar K^{\ast0} 
(\to \pi^+ K^-)]}
{d \cos \theta~d \varphi~d \cos \psi}= 
\frac{9}{64 \pi}\left[\cosh(\Delta\Gamma_st/2)-\cos(\Delta M_st)\right]
e^{-\Gamma_s t}}\nonumber\\
&&\times\left\{ f_1 |A_0|^2  + f_2 |A_{\|}|^2 + f_3 |A_{\perp}|^2 
 - f_4\, {\mbox  Im}\,(A_{\|}^{\ast} A_{\perp})  
 + f_5\, {\mbox Re}\,(A_0^{\ast} A_{\|})
 + f_6\, {\mbox Im}\,(A_0^{\ast} A_{\perp})\right\}.\nonumber\\
\end{eqnarray}
In the case of $\Delta\Gamma_s \, \to \,0$, we have
\begin{equation}
\cosh(\Delta\Gamma_st/2)+\cos(\Delta M_st) \, \to \,
2\cos^2(\Delta M_st/2),
\end{equation}
\begin{equation}
\cosh(\Delta\Gamma_st/2)-\cos(\Delta M_st) \, \to \,
2\sin^2(\Delta M_st/2).
\end{equation}
Consequently, the expressions listed above reduce to those given
in Ref.~\cite{DDF} for the flavour-specific 
$B_d \to J/\psi[\to \ell^+ \ell^-] K^{\ast}[\to K^\pm \pi^\mp]$ modes with 
the assumption of $|A_f|=|\bar A_f|$.

\newpage

\end{document}